\documentclass[12pt,twoside]{article}
\usepackage[T1]{fontenc}
\usepackage[utf8]{inputenc}

\usepackage{geometry}
\usepackage{fancyhdr}
\usepackage{authblk}

\usepackage{amsmath}
\usepackage{amsfonts}
\usepackage{amssymb}
\usepackage{siunitx}
\usepackage{xcolor}

\usepackage{graphicx}
\usepackage[
  format=default,
  font={small},
  labelfont={bf},
  belowskip=5pt
]{caption}

\usepackage{tabularx}
\usepackage{booktabs}

\usepackage[authoryear]{natbib}
\usepackage[
  colorlinks = true,
  linkcolor = magenta,
  citecolor = magenta,
  urlcolor = magenta
]{hyperref}

\title{RainScaleGAN: a Conditional Generative Adversarial Network for Rainfall Downscaling}

\author[a]{\normalsize Marcello Iotti\thanks{Corresponding author: Marcello Iotti, \href{mailto:marcello.iotti@ge.imati.cnr.it}{marcello.iotti@ge.imati.cnr.it}}}
\author[b]{\normalsize Paolo Davini}
\author[c]{\normalsize Jost von Hardenberg}
\author[d]{\normalsize Giuseppe Zappa}

\affil[a]{\footnotesize Institute for Applied Mathematics and Information Technologies, CNR, Genoa, Italy}
\affil[b]{\footnotesize Institute of Atmospheric Sciences and Climate, CNR, Turin, Italy}
\affil[c]{\footnotesize Department of Environment, Land and Infrastructure Engineering, Politecnico di Torino, Turin, Italy}
\affil[d]{\footnotesize Institute of Atmospheric Sciences and Climate, CNR, Bologna, Italy}

\date{}

\begin{document}

\noindent \textsl{\textcolor{gray}{This Work has been submitted to Artificial Intelligence for the Earth Systems (AIES).
Copyright in this Work may be transferred without further notice.}}

{\let\newpage\relax\maketitle}

\begin{abstract}
To this day, accurately simulating local-scale precipitation and reliably reproducing its distribution remains a challenging task.
The limited horizontal resolution of Global Climate Models is among the primary factors undermining their skill in this context.
The physical mechanisms driving the onset and development of precipitation, especially in extreme events, operate at spatio-temporal scales smaller than those numerically resolved, thus struggling to be captured accurately.

In order to circumvent this limitation, several downscaling approaches have been developed over the last decades to address the discrepancy between the spatial resolution of models output and the resolution required by local-scale applications.

In this paper, we introduce RainScaleGAN, a conditional deep convolutional Generative Adversarial Network (GAN) for precipitation downscaling.
GANs have been effectively used in image super-resolution, an approach highly relevant for downscaling tasks.
RainScaleGAN's capabilities are tested in a \textit{perfect-model} setup, where the spatial resolution of a precipitation dataset is artificially degraded from \num{0.25}\si{\degree}$\times$\num{0.25}\si{\degree} to \num{2}\si{\degree}$\times$\num{2}\si{\degree}, and RainScaleGAN is used to restore it.
The developed model outperforms one of the leading precipitation downscaling method found in the literature.
RainScaleGAN not only generates a synthetic dataset featuring plausible high-resolution spatial patterns and intensities, but also produces a precipitation distribution with statistics closely mirroring those of the ground-truth dataset.
Given that RainScaleGAN's approach is agnostic with respect to the underlying physics, the method has the potential to be applied to other physical variables such as surface winds or temperature.

\end{abstract}

\newpage

\vspace{1em}
\noindent \textbf{\small Significance Statement}  
\vspace{1em}

\noindent {\small Accurately predicting local precipitation is difficult due to the limitations of current climate models.
These models struggle to capture the small-scale processes causing precipitation, especially those leading to extreme events.
To address this, we developed a new tool that uses advanced Artificial Intelligence techniques to improve rainfall predictions.
This tool takes low-resolution precipitation data and enhances it to high resolution, providing more detailed rainfall patterns.
Our results show that the tool performs better than one of the leading existing methods.
This advancement could lead to more precise climate projections, better preparation for extreme weather, and suggests further exploration on additional weather variables.}  

\bigskip


\section{Introduction}
Global Climate Models (GCMs) are nowadays the primary tools for the investigation of the climate system, its mechanisms and its changes.
Despite the remarkable skill achieved in the recent decades across a wide range of applications, and their continuous evolution, they still lack accuracy in the reproduction of the precipitation distribution \citep[cf.][]{Sha2020DeepLearningBasedGriddedDownscaling}.
The most relevant reason for such limitation can be traced back to the spatial resolution at which most of the current state-of-the-art GCMs are run, usually falling within the range \num{50}-\num{200} \si{\kilo\meter}.
Such horizontal resolution, much coarser than the typical spatial scale of precipitation and convective structures, can reasonably capture the synoptic and part of the meso-scale atmospheric circulation, but is too coarse to accurately represent smaller-scale phenomena, particularly where proper modeling requires a precise representation of surface meteorological variables on topographically complex terrain.
Weather and climate models relies on specific \textit{parameterisations} to tackle this inadequacy, and despite notable improvements in recent years this approach still presents limitations.
In particular, numerical models suffer from an imperfect physical representation of precipitation: they usually simulate convective and stratiform precipitation independently, resulting in an inaccurate precipitation distribution, where typically the occurrence of light rain (drizzle) is overestimated, while dry days and high to extreme events are underestimated \citep[see e.g.][]{Piani2010Statisticalbiascorrection}.

Beyond simulating the atmosphere for research purposes, atmospheric modeling serves societally-relevant goals.
It plays a crucial role in supporting a range of applications, including hydrological modeling, water management and agriculture.
In broader terms, it enables scientifical evidence-based decision-making processes for policymakers, engineers and planners, who need to understand and formulate a response to predicted events.
All these applications are highly sensitive to the precipitation input they receive, to its resolution and to the details of its fine-scale distribution, thus requiring greater accuracy and a finer spatial distribution than what GCMs provide.

In atmospheric sciences, the term \textit{downscaling} refers to any operation aimed at inferring high-resolution variables from lower-resolution data.
Many techniques exist, founded on the assumption that the large-scale configuration of the atmosphere strongly influences variables at the local scale.
Downscaling techniques can be classified into two main groups, each approaching the task differently \citep[for an in-depth review see][]{Maraun2010Precipitationdownscalingclimate}.
On the one hand, \textit{dynamic downscaling} uses higher-resolution \textit{Regional Climate Models} nested within lower-resolution GCMs, which provide boundary conditions to them \citep{Feser2011RegionalClimateModels,Rummukainen2010Stateoftheartregionalclimate}.
While these models have a strong physical basis, they come with large computational costs, limiting their coverage to specific areas and a restricted number of simulations.
On the other hand, \textit{statistical downscaling} is a post-processing technique that establishes statistical relationships between large-scale predictors and small-scale predictands \citep{Wilby1997Downscalinggeneralcirculation,Rummukainen1997Methodsstatisticaldownscaling,Wilby1999comparisondownscaledraw,Dibike2005Hydrologicimpactclimate}.
Methods within this category are computationally less expensive, yet their calibration relies on high-quality local-scale data, which may not always be available everywhere.
Moreover they might not be easily transferable to different regions of the globe.
A particular category of statistical methods are the so-called \textit{stochastic downscaling} methods, a form of weather generator \citep{Maraun2010Precipitationdownscalingclimate} which, starting only from large-scale precipitation fields, can generate fine-scale downscaled fields with a realistic spatial correlation structure and amplitude distribution \citep[for a comparison see for example][]{Ferraris2003comparisonstochasticmodels}.

In recent years, applying Machine Learning (ML) techniques originally developed in image processing to downscaling tasks has produced remarkable results.
The exploration of such techniques in a context different from their origin, has been driven by the similarity between downscaling and the so-called image super-resolution (upsampling)\footnote{Note that, somewhat confusingly, the term \textit{upsampling} (\textit{downsampling}) in the field of image processing refers to the process of increasing (decreasing) the resolution of an image. This is exactly the opposite of the terms \textit{upscaling}/\textit{downscaling} used in meteorology.}, which is the process of enhancing the resolution of an image \citep{Reichstein2019Deeplearningprocess}.
To date, the most successful ML models in the field of image processing are based on Convolutional Neural Networks (CNNs), leading many authors to tackle the downscaling problem using CNNs \citep{Sha2020DeepLearningBasedGriddedDownscaling,Kumar2021Deeplearningbased,Wang2021DeepLearningDaily}.
Further advancements have come from applying Generative Adversarial Networks (GANs) \citep{Goodfellow2014GenerativeAdversarialNets,Goodfellow2020Generativeadversarialnetworks} to the downscaling problem.
The goal of GANs is to train a neural network, called \textit{generator}, to generate examples that mimic the probability distribution of the training data.
A complementary neural network, the \textit{discriminator}, is designed to assess the generated examples, distinguishing them from the training data, and then encouraging the generator to enhance its performance.
\citet{Ledig2017PhotoRealisticSingleImage} applied GANs to image super-resolution, while \citet{Leinonen2021StochasticSuperResolutionDownscaling} introduced a recurrent super-resolution GAN able to generate ensembles of plausible high-resolution atmospheric fields from their low-resolution (upscaled) counterparts.
\citet{Ravuri2021Skilfulprecipitationnowcasting} addressed the problem of the so-called \textit{nowcasting}, developing a deep generative model for the probabilistic short-term prediction of radar-measured precipitation.
\citet{Harris2022GenerativeDeepLearning} and \citet{Price2022Increasingaccuracyresolution} extended the problem addressed by Leinonen, building models mapping from multiple low resolution atmospheric fields (including precipitation) from a numerical weather prediction model, to high resolution radar-measured precipitation.
More recently, \citet{Annau2023AlgorithmicHallucinationsNearSurface} developed a super-resolution GAN-based model trained on non-idealized pairs consisting of low-resolution (80 km) reanalysis data and 10-m wind component fields from a convection-permitting (4 km) model driven by the same low-resolution dataset.
Their model aims to reproduce fine-scale details consistent with the convection-permitting simulation, effectively capturing its internal variability.

In this paper, we will demonstrate how a GAN with a simple architecture can effectively downscale precipitation.
By relying solely on the low-resolution precipitation field as a predictor, our approach can be easily generalized to any part of the globe, as it is independent of explicitly incorporating the topographic features of the geographical region under investigation, nor does it require additional external sources of information.
However, it is important to note that extending the method to other regions would require additional training with region-specific precipitation data.
We will conduct the training and testing of the model in the so-called \textit{perfect model setup} \citep[pure super-resolution, cf.][]{Harris2022GenerativeDeepLearning}, reducing the resolution of training data through spatial aggregation, and using our model to restore the lost original resolution.
We will demonstrate how the generated dataset closely mirrors the statistical properties of the original dataset.
Additionally, our trained generator proves to be more effective in producing high-resolution precipitation fields compared to RainFARM \citep{Rebora2006RainFARMRainfallDownscaling,DOnofrio2014StochasticRainfallDownscaling,Terzago2018Stochasticdownscalingprecipitation}, a state-of-the art stochastic downscaling method.

The structure of the paper is as follows: Section \ref{sec:data} presents the data used in our experiments and their pre-processing.
In Section \ref{sec:methods}, we define the task we addressed, describe the model architecture, outline the training process, list the metrics used to assess the performance of the model, and briefly introduce RainFARM, the alternative downscaling method used as a baseline to assess RainScaleGAN's skills.
The next Section \ref{sec:results} presents the results of the experiments, describing the training process, model validation and testing, with the final comparison with RainFARM.
The final Section \ref{sec:discussion_conclusion} discusses the results, the limitations of the adopted framework, and possible future developments.


\section{Data}
\label{sec:data}

The ERA5 \citep{Hersbach2023ERA5hourlydata} reanalysis data for total precipitation has been used throughout the entire Machine Learning exercise, and for conducting the downscaling process using RainFARM.
This variable represents the cumulative amount of liquid and solid precipitation, resulting from both large-scale and convective precipitation.
The spatial covering is global, with a resolution of \num{0.25}\si{\degree}$\times$\num{0.25}\si{\degree}, and the temporal resolution equals \num{1} hour.

To demonstrate our model we chose a region of interest centred on the Alpine arch, spanning latitudes 38N to 53.75N and longitudes 3E to 18.75E.
This region includes both sea and land for the majority of the Italian Peninsula, Austria, the Czech Republic, the central and southern part of Germany, Switzerland, the Netherlands, Belgium, the eastern part of France, and some portions of the neighbouring countries.
Such a choice is rather arbitrary, but our model is designed to be agnostic regarding the region to which it is applied.
Additionally, the selected region encompasses a topographically complex terrain, due to the presence of orography, making it a suitable testbed for a rainfall downscaling technique.

\subsection{Data Source and Preprocessing}
\label{subsec:data_source_processing}

ERA5 total precipitation has been obtained through the Copernicus Climate Data Store (CDS) \citep{CopernicusClimateChangeServiceC3S2023ERA5hourlydata}.
In this archive, ERA5 data is interpolated onto a regular latitude-longitude grid.
The total precipitation is derived from short (18-hour) forecasts, run twice a day from the 06 and 18 UTC analyses.
The accumulation is carried out for the hour ending at the date and time of validity.
We computed the daily precipitation by taking the average of the hourly values within the same date, and then multiplying by 24 (the number of hours in a day).
This value is not coincident with the precipitation actually accumulated during the corresponding 24 hours, as the precipitation with valid time 00:00 UTC is accumulated from 23:00 to 23:59 UTC of the previous day.
In the present study we do not plan a comparison with measured data, therefore such inconsistency is not relevant.

The data related to the domain of interest has been extracted, without performing any further spatial interpolation, resulting in precipitation fields of $64\times64$ grid points (a box of approximately  $1800\times1300$ \si{\kilo\metre}).
A spatial filter has been applied to the dataset to exclude days with extremely low precipitation, as a measure to counterbalance part of the drizzle problem.
Additionally, we observed that exposing the GAN to non-meaningful samples slows down its convergence.
The filter is implemented by computing the spatial average of the precipitation across the entire domain for each sample of the dataset.
Days are excluded from the dataset if this quantity falls below a small threshold, arbitrarily set at \num{1} \si{\milli\meter}.
The resulting two-dimensional daily precipitation fields constitute the \textit{examples} used for training the GAN.

Feature scaling is a fundamental preprocessing step in most ML tasks.
It enhances the convergence speed and performance of ML optimisation algorithms, effectively preventing gradient descent issues.
Moreover, it helps in handling skewed data and reducing the impact of outliers, balancing the influence of features.
Therefore, the following transformations are applied to the original precipitation rate $x$, expressed in \si{\milli\meter}/day:
\begin{enumerate}
    \item Square root transformation $\sqrt{x}$.
    \item Rescaling (min-max normalisation), applied \textit{separately} to each grid point according to the following formula:
    \begin{equation}\label{eq:GAN_scaling}
    x_{\text{scaled}} = m + \frac{x - x_{\text{min}}}{x_{\text{max}}-x_{\text{min}}} (M - m)
    \end{equation}
    where $x_{\text{min}}$ and $x_{\text{max}}$ are the minimum and the maximum values of the time series for that grid point, and  the chosen feature range is $[m,M]=[-1,1]$.
\end{enumerate}
As precipitation has a strongly positively skewed distribution, the reexpression \citep[see e.g.][]{Wilks2011StatisticalMethodsAtmospheric} using the square root transformation contributes to obtaining a more symmetric distribution, facilitating the analysis of data and improving the performance of the ML model.
Additionally, it avoids the issue of zeros related to the commonly used logarithmic transformations.
Regarding the latter transformation (min-max scaling), since the value of precipitation over each grid point can be considered as a feature influenced by the underlying topography and specific precipitation-generating processes, the rescaling ensures a consistent treatment of the entire precipitation field under consideration.

\subsection{Data Subsets}
Data selection is critical in constructing a data-driven model.
Proper sampling of predictor variability is essential for achieving model generalizability.
In ML practice, it is standard to divide the data into training, validation, and test subsets, ensuring that these subsets are independent of each other.

However, meteorological data can be conceptualized as time series that are intrinsically autocorrelated over finite spatial and temporal domains.
Therefore, the standard procedure often used in deep learning research - extracting random samples from the available data, assuming each instance is independent, and arbitrarily assigning them to training, validation, and test sets - is inappropriate.
This approach can overestimate the skill of the model because random sampling introduces correlations among the three subsets, thereby incorporating information into the test set that has already been used in training.
To address this, we adopt the strategy of random block sampling \citep{Schultz2021Candeeplearning}, where the dataset is split into blocks with durations much greater than the period of time considered to contribute the most to autocorrelation (a few days).
A downside of this method is that it assumes there are no significant long-term trends in the distribution of precipitation, which contrasts with the effects of climate change.
Nevertheless, it can serve as a useful indicator of the effectiveness of the ML model in handling such statistical changes over time, assessing its robustness and applicability, for example, in the context of climate projections.

The ERA5 dataset we downloaded, spanning the years from 1940 to 2022, has been divided into three consecutive subsets: data from 1940 to 1998 is used as the training set (15692 examples after filtering), data from 1999 to 2010 is used as the validation set (3261 examples), and data from 2011 to 2022 is used as the test set (3134 examples).
To apply minibatch stochastic gradient descent \citep{Goodfellow2016DeepLearning}, shuffling is applied on each of these three subsets.
This shuffling aims to ensure independence between the examples within each minibatch, as well as between the minibatches themselves.


\section{Methods}
\label{sec:methods}

\subsection{Definition of the task}
\label{subsec:definition_task}

\begin{figure}[!t]
	\centering
	\includegraphics[width=0.8\textwidth]{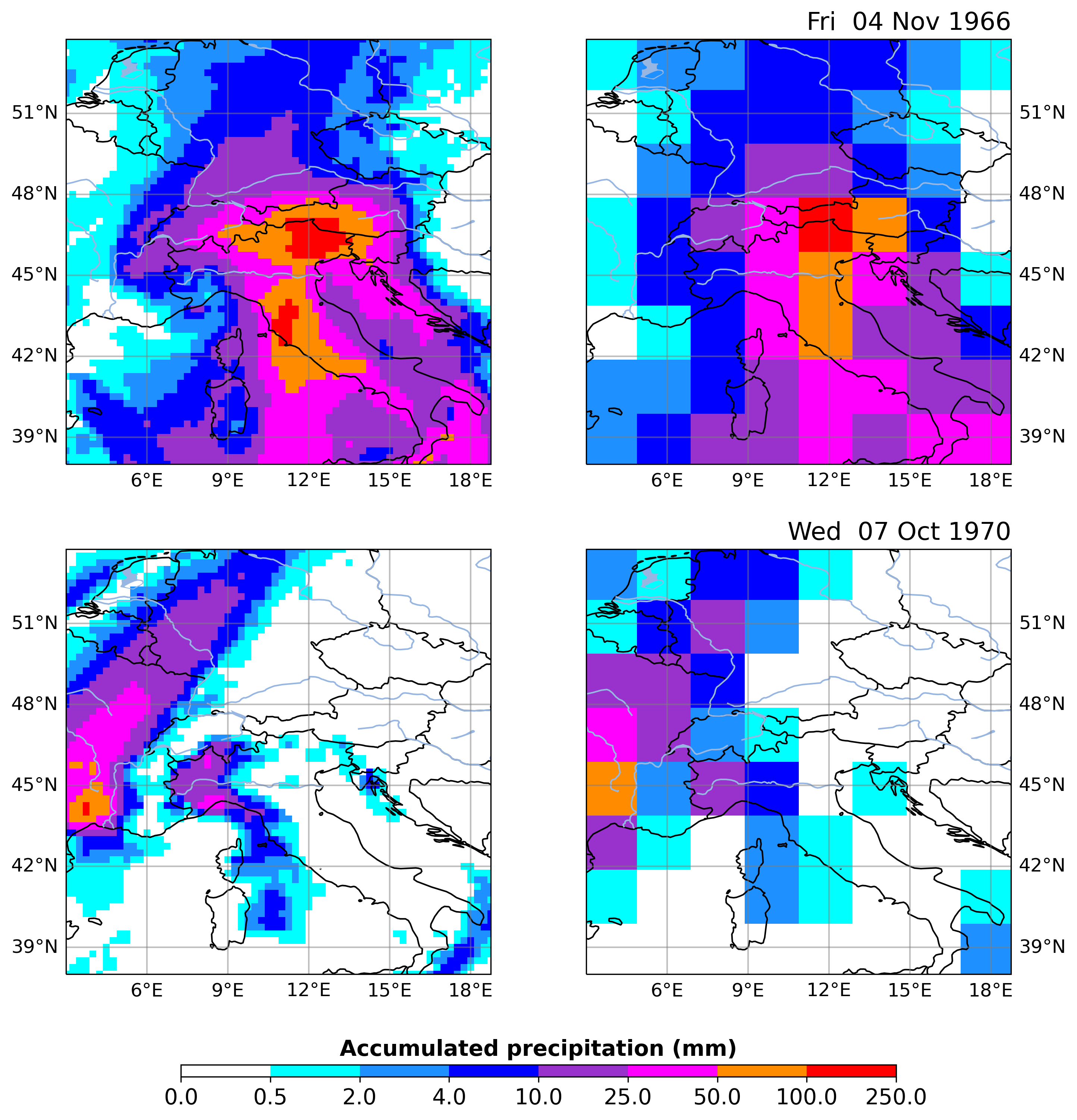}
	\caption{ERA5 daily total accumulated precipitation (left) and corresponding coarsened version (right) for two sample days (04 November 1966 and 07 October 1970).
	The original ERA5 examples have a spatial resolution of \num{0.25}\si{\degree}$\times$\num{0.25}\si{\degree} and consist of 64x64 grid points.
	The coarsened versions, obtained with an upscaling factor of 8, have a spatial resolution of \num{2}\si{\degree}$\times$\num{2}\si{\degree} and consist of 8x8 grid points.}
	\label{fig:ERA5_coarse}
\end{figure}

An inherent challenge in evaluating a new downscaling technique is the unavoidable presence of biases between the predictor and target datasets.
In order to circumvent this issue, we adopt what in literature is referred to as a \textit{perfect-model} setup \citep{Terzago2018Stochasticdownscalingprecipitation}.
This involves taking a high-resolution precipitation dataset, measured or simulated, and artificially degrading its spatial resolution through an aggregation operation.
The downscaling model is then tasked with restoring the lost resolution based on this smoother, low-resolution precipitation field.
This allows for the assessment of the skill of the downscaling method, measuring whether the produced field reflects the correct rainfall patterns and statistical properties of the true field \citep{Rebora2006RainFARMRainfallDownscaling}.
In terms closer to those used in the machine learning field, it is also called \textit{pure super-resolution} \citep{Harris2022GenerativeDeepLearning}.

In order to apply this procedure to our study, we performed a spatial aggregation (upscaling) on the ERA5 daily total precipitation, reducing its spatial resolution by a factor of 8, from \num{0.25}\si{\degree}$\times$\num{0.25}\si{\degree} to \num{2}\si{\degree}$\times$\num{2}\si{\degree}.
The operation consists in taking the average of precipitation across groups of $8\times 8$ adjacent grid cells.
The resulting coarsened field covers an area of $8\times 8$ grid cells.
Figure \ref{fig:ERA5_coarse} displays some examples of the outcome of the described operation.

\subsection{Model architecture}
\label{subsec:model_architecture}

\begin{figure}[!t]
	\centering
    \includegraphics[width=0.75\textwidth]{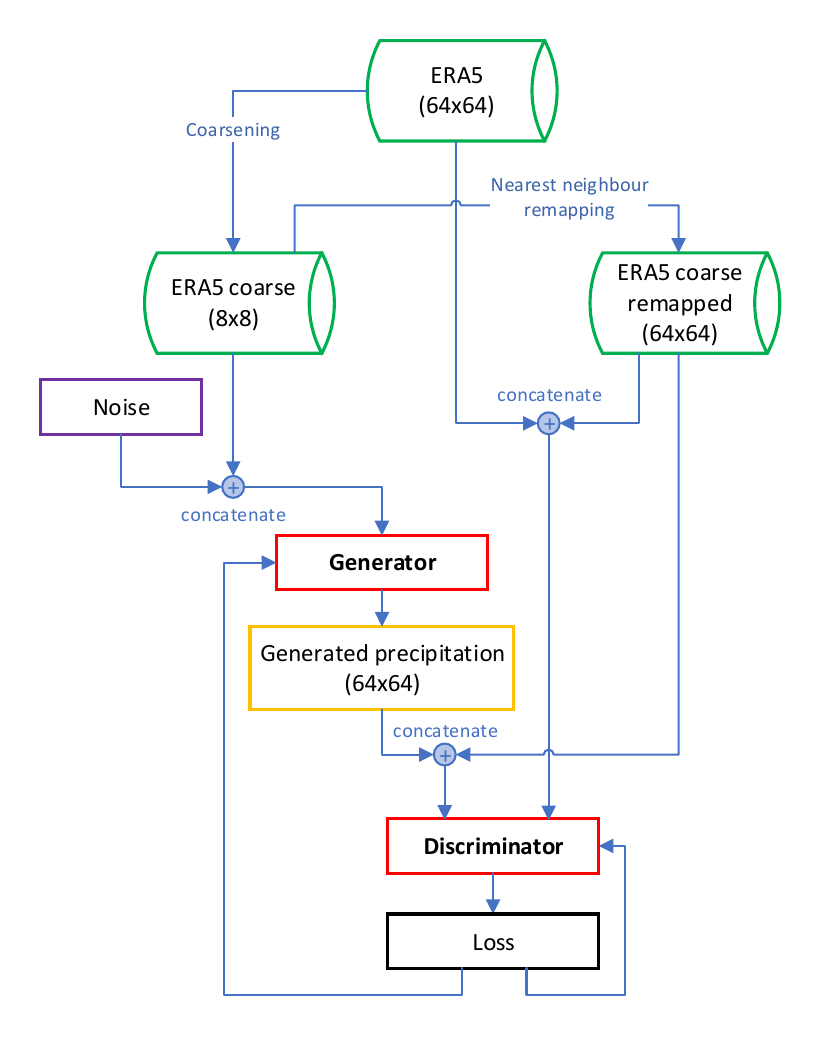}
	\caption{Information flow during the model training.}
	\label{fig:GAN_schematic}
\end{figure}

The model we constructed is a conditional GAN \citep{Mirza2014ConditionalGenerativeAdversarial} i.e.~a GAN in which both the generator and discriminator receive, as additional input, conditioning data aimed at directing the generation process.
In our case, this conditioning data consist of the low resolution version of the daily precipitation field to be downscaled.
This source of information is conditioning in the sense that it provides the large-scale structure of the precipitation field that the generated fine-scale example is required to adhere to.
The task of the generator is to produce a precipitation field at the target spatial resolution, using an input composed of:
\begin{itemize}
	\item The corresponding low-resolution conditioning field.
	\item A source of noise, in this case an array of random numbers drawn from a normal distribution with mean \num{0} and standard deviation \num{0.02}.
\end{itemize}
The noise source in the generator implies that it is capable of producing an indefinite number of examples consistent with the structure of low-resolution conditioning field.
The task of the discriminator is to distinguish the predictions of the generator from the corresponding ``ground-truth'' fields from the training set.
The discriminator is fed with either ground-truth or generated examples, each one together with the corresponding low-resolution precipitation field, always drawn from the upscaled version of the training data.
Please note that the upscaled counterparts of the generated precipitation are never used.
Due to the architecture of the neural networks we implemented for the two components of RainScaleGAN, inputs must be concatenated.
In the case of the discriminator, this operation requires that input fields have the same dimension.
We thus performed a nearest-neighbour remapping on the low-resolution dataset, generating a rainfall field with the same information content, but with spatial resolution matching the one of the target, high-resolution dataset.
The output of the discriminator is used to calculate the loss function for both the discriminator itself and the generator.
This way, the discriminator guides the training process, providing a feedback to the generator, ideally enabling it to improve its performance during the training process.
Figure \ref{fig:GAN_schematic} presents an overview of the described process, illustrating the interplay between information sources, models, and their outputs during the training phase.

\begin{figure}[!t]
	\centering
	\includegraphics[width=0.75\textwidth]{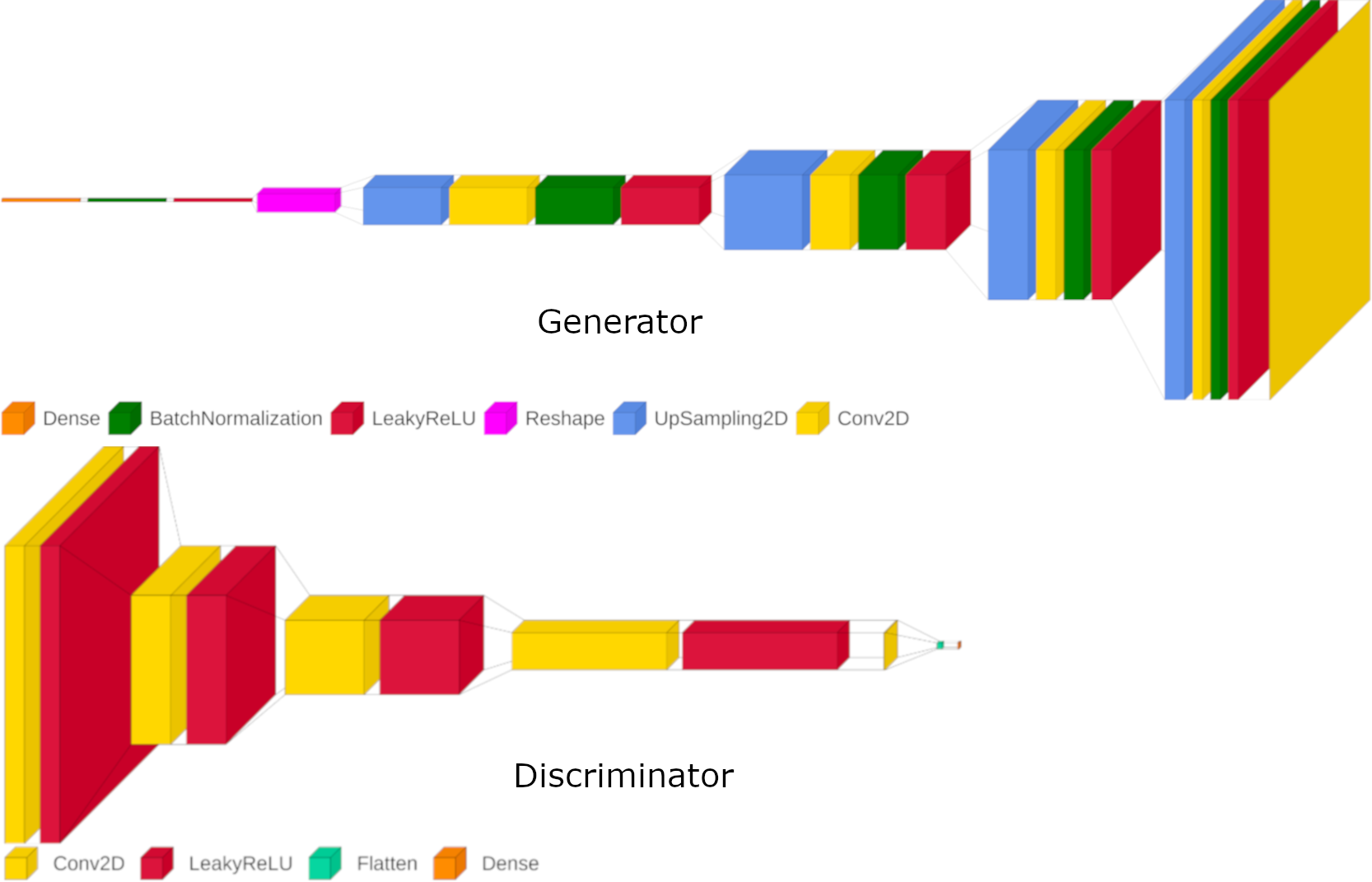}
	\caption{The architecture of the networks composing RainScaleGAN.
	The input layers of the two networks (coarse image + noise source for the generator, generated/ground-truth image + corresponding coarse image for the discriminator) are not shown.
	(Top) Generator architecture. The upsampling layers have upsampling factors of (2,2), thereby doubling the number of rows and columns of their input.
	The intermediate convolutional layers have a number of kernels equal to 256, 128, 64, and 32, respectively.
	The final convolutional layer has a single kernel and is activated with the hyperbolic tangent function.
	(Bottom) Discriminator architecture.
	The convolutional layers, except the last one, have a number of kernels equal to 64, 128, 256, 512, respectively, and strides (2,2).
	Each of them halves the height and width of the field it receives as input.}
	\label{fig:gan_architecture}
\end{figure}

Figure \ref{fig:gan_architecture} illustrates the architecture of the generator and discriminator, both implemented as deep convolutional artificial neural networks.
The input to the generator is a 1-dimensional array constructed by flattening the low-resolution input field to be downscaled and concatenating it with the array of random numbers mentioned above.
This input is initially mapped through a dense layer to a 3-dimensional tensor of appropriate size, depending on the number of grid points in the precipitation field to be generated.
Following there are four blocks of layers designed to map this tensor to the target field.
Such number of blocks depends on the spatial extent of the target rainfall field and, ultimately, on the upscaling factor i.e.~the ratio between the spatial resolutions of the low- and high-resolution fields.
The essential components of these blocks consist of a 2-dimensional upsampling layer followed by a 2-dimensional convolutional layer, with increasing horizontal dimensions (height and width) and decreasing number of filters.
Specifically, the upsampling layers have upsampling factors of (2,2), doubling the height and width of the tensor they receive, while the convolutional layers have a stride of 1 in both directions and number of filters equal to 256, 128, 64, 32, respectively.
Batch normalization \citep{Ioffe2015BatchNormalizationAccelerating} and Leaky Rectified Linear Unit (ReLU) activation with a negative slope of 0.2 are applied at the end of both the input block and each of the aforementioned convolutional blocks.
The network concludes with a 2-dimensional convolutional layer with a single filter, employing the hyperbolic tangent as activation.
This layer aims to generate the final image corresponding to the rainfall field at the target resolution.

The structure of the discriminator mirrors that of the generator.
It takes as input the pair of ground-truth/generated high-resolution rainfall fields and their corresponding low-resolution rainfall field which - as mentioned above - has been remapped with the nearest-neighbour method.
These fields are concatenated along their depth (i.e.~the images are stacked) and then passed to the discriminator.
Four blocks consisting of convolutional layers with strides (2,2), to reduce the height and width of the input by half at each layer, and increasing number of filters (64, 128, 256, 512) are employed.
Each block is activated using the Leaky ReLU function with a negative slope of 0.2.
The network concludes with a single-filter convolutional layer, densely connected to a single unit, with linear activation (i.e.~no final activation is applied).

The size of the convolutional filters, both in the generator and the discriminator, is a parameter that we optimized during the validation phase.
The optimal generator and discriminator that we selected have approximately 4.3 million and 2.8 million trainable parameters, respectively (compare with the training set size: 15692 examples, each consisting of a $64\times64$ grid points precipitation field).

\subsection{Training}
Training a GAN involves the simultaneous training of two models: the discriminator $D$, which aims to maximise the probability that it assigns the correct label to real (from the training set) and fake (from the generator) examples, and training the generator $G$, which aims to maximise the probability that $D$ mistakenly assigns the label ``real'' to generated examples.
In other words, the training is a minimax game with a value function $V(G,D)$ \citep{Goodfellow2014GenerativeAdversarialNets}:
\begin{equation}
\label{eq:GAN_value}
\begin{split}
    \min_G \max_D V(D, G) = &\; \mathbb{E}_{\mathbf{x} \sim p_{\text{data}}(\mathbf{x})}[\log{D(\mathbf{x})}] +\\
    & + \mathbb{E}_{\mathbf{z} \sim p_{\mathbf{z}}(\mathbf{z})}[\log{(1 - D(G(\mathbf{z})))}]
\end{split}
\end{equation}
where $p_{\text{data}}$ and $p_{\mathbf{z}}$ are the distribution of training data $\mathbf{x}$ and of a noise variable $\mathbf{z}$, respectively.
Under appropriate assumptions \citep[cf.][]{Goodfellow2014GenerativeAdversarialNets}, the minimax game expressed by Equation \ref{eq:GAN_value} translates into minimising the Jensen-Shannon divergence\footnote{The Jensen–Shannon divergence is a measure of the similarity between two probability distributions.} between the distribution of training data and that of generated data.
This divergence suffers from the problem of not being continuous with respect to the generator parameters \citep{Arjovsky2017WassersteinGAN} and its minisation often leads to discriminator saturation with resulting vanishing gradients \citep{Gulrajani2017ImprovedTrainingWasserstein}.
Wasserstein GANs (WGAN) \citep{Arjovsky2017WassersteinGAN} are designed to address these issues.
The training objective of a WGAN is expressed by:
\begin{equation}
	\min_G \max_{D\in\mathcal{D}}
	\mathbb{E}_{\mathbf{x} \sim p_{\text{data}}} [D(\mathbf{x})] - \mathbb{E}_{\mathbf{z} \sim p_{\mathbf{z}}(\mathbf{z})} [D(G(\mathbf{z}))]
\end{equation}
where $\mathcal{D}$ is the set of 1-Lipschitz function. 
Optimising the discriminator (referred to as the \textit{critic} in the foundational paper), the minimisation of the previous value function with respect to the generator minimises the Earth-Mover (Wasserstein-1) distance:
\begin{equation}
	W(p,q) = \inf_{\gamma \in \Pi(p,q)} \mathbb{E}_{(x, y) \sim \gamma} [\|x - y\|],
\end{equation}
being $\Pi(p,q)$ the set of joint distributions $\gamma(x,y)$ with marginals are $p$ and $q$, respectively.
The value function based on this distance exhibits better properties than the original value function, making the optimization of the generator simpler.
Following \citet{Gulrajani2017ImprovedTrainingWasserstein}, we enforce the Lipschitz constraint on $D$ by introducing a gradient penalty term in the discriminator loss.

The GAN framework extends to a conditional model by incorporating auxiliary information $\mathbf{y}$ into both the generator and discriminator \citep{Mirza2014ConditionalGenerativeAdversarial}.
In a downscaling task, this auxiliary information consists of the low-resolution field to be refined.
Thus, the objective functions for the generator and discriminator of the conditional Wasserstein GAN with gradient penalty (WGAN-GP) used in this study are:
\begin{align}
\begin{split}
\label{eq:GAN_loss_discriminator}
    L_D = &\; \mathbb{E}_{\mathbf{x} \sim p_{\text{data}}} [D(\mathbf{x}|\mathbf{y})] +\\
    & -
	\mathbb{E}_{\mathbf{z} \sim p_{\mathbf{z}}(\mathbf{z})} [D(G(\mathbf{z|\mathbf{y}})|\mathbf{y})] + \\
	& +\lambda
	\left( \mathbb{E}_{\hat{\mathbf{x}} \sim \hat{P}_{\mathbf{x}}} \left[ \left(\|\nabla_{\hat{\mathbf{x}}}D(\hat{\mathbf{x}})\|_2 - 1\right)^2 \right] \right) \\
\end{split}
\\[2ex]
\begin{split}
\label{eq:GAN_loss_generator}
    L_G = &\; \mathbb{E}_{\mathbf{z} \sim p_{\mathbf{z}}(\mathbf{z})} [D(G(\mathbf{z|\mathbf{y}})|\mathbf{y})] \\
\end{split}
\end{align}
being $\lambda$ a constant representing the weight of the gradient penalty term, set equal to 10 in accordance with \citet{Gulrajani2017ImprovedTrainingWasserstein}, and the samples $\hat{\mathbf{x}}$ are defined by:
\begin{equation}
	\hat{\mathbf{x}} = \epsilon \mathbf{x} + (1 - \epsilon) G(\mathbf{z})
\end{equation}
where $\epsilon$ is a random number drawn from a uniform distribution $U[0, 1]$.
Both the generator and the discriminator aim to maximise their respective objective functions defined as above.
Unlike some other conditional GAN approaches for downscaling, we did not implement a content loss term.
The relevance of content loss is evident in studies that involve observations \citep[e.g.][]{Harris2022GenerativeDeepLearning} or aim to emulate specific characteristics of the process generating the target dataset \citep[e.g.][]{Annau2023AlgorithmicHallucinationsNearSurface}, often with a stronger emphasis on improving per-grid-cell metrics.
In contrast, its role appears to be less prominent in perfect-model setups like ours \citep[cf.][]{Leinonen2021StochasticSuperResolutionDownscaling}.
Additionally, this choice allows us to assess the impact of the objective function of the generator, as defined in Equation \ref{eq:GAN_loss_generator} - commonly referred to as the \textit{adversarial component} - on effectively guiding the super-resolution task as formulated in this study, independent of any content loss.

Following \citet{Arjovsky2017WassersteinGAN}, during the training cycle of RainScaleGAN, we alternate between 5 iterations of training for the discriminator and 1 iteration of training for the generator.
The Adam optimiser \citep{Kingma2017AdamMethodStochastic} with a learning rate of \num{2e-4} has been chosen for both the neural networks.

\subsection{Skill metrics}
\label{subsec:skill_metrics}

To assess the performance of the model, as well as to monitor training and conduct validation, we employed the following set of metrics.

As a simple indicator of the quality of the generated precipitation fields, useful for evaluating the convergence of the training process, we use the root-mean-square error:
\begin{equation}
	RMSE = \sqrt{\frac{1}{N} \sum_{i=1}^{N} \left( x_{\text{true},i} - x_{\text{gen},i} \right)^2}
\end{equation}
In order to have a more precise evidence of the ability of the GAN to reconstruct the spatial structure and variability of the generated rainfall field, we calculate the log spectral distance (LSD) between the spatial radial spectrum of the generated dataset and the corresponding spectrum of the ERA5 subset:
\begin{equation}
	LSD = \sqrt{\frac{1}{N} \sum_{i=1}^{N} \left( 10 \log_{10}{\frac{P_{\text{true},i}}{P_{\text{gen},i}}} \right)^2}.
\end{equation}
The power spectra $P_{\text{true}}$ and $P_{\text{gen}}$ are obtained by performing the Fourier transform in the physical (2-dimensional) space of the precipitation field, averaging along the time axis over all the examples within the dataset in question.
Then, a binned average is applied in the $k$-space, each bin being centred on each of the discrete Fourier wavenumbers that can be defined in the physical space.
This operation is equivalent to collapsing over all angular directions the 2-dimensional spectrum, obtaining a 1-dimensional spectrum \citep[cf.][]{Harris2022GenerativeDeepLearning}.

The two metrics presented above (henceforth collectively referred to as \textit{image metrics}) are borrowed from the practice of image processing.
Although they constitute an important reference point for evaluating the model's skill, they lack in describing the statistical, and in some sense physical, properties of the generated dataset.
To provide a more comprehensive assessment of these properties, considering their importance especially within the context of climate studies, we extended our suite of metrics to include a set of basic statistics for the generated dataset.
These statistics were not only monitored during training, but were also crucial for model selection and in the final assessment of the trained model, facilitating a comparison with the alternative downscaling method.
In the following, we will collectively refer to these metrics as \textit{statistical metrics}.
Specifically, we considered the climatology and the standard deviation of the time series for daily total precipitation, calculated grid point-wise:
\begin{align}
	\text{Clim}(i,j) &= \frac{\sum_t x_t^{(i,j)}}{T} \\
	\text{StD}(i,j) &= \sqrt{ \frac{\sum_t \left( x_t^{(i,j)} - \text{Clim}(i,j) \right)^2}{T} }
\end{align} 
where $x_t^{(i,j)}$ is the amount of the daily total accumulated precipitation on the grid point $(i,j)$ for the day $t$, while $T$ is the total number of days in the dataset.
Moreover, we compute the 95\textsuperscript{th} and 99\textsuperscript{th} percentiles of daily total precipitation, once again on a grid point-wise basis.
The climatology and standard deviation enable the assessment of the mean statistical properties of the generated dataset.
The uppermost percentiles are important in evaluating the GAN's capability to accurately capture both the magnitude and the localisation of extreme events.

\subsection{Alternative method}
\label{subsec:alternative_method}

As a baseline for comparing the performance of the constructed model, we chose RainFARM \citep{Rebora2006RainFARMRainfallDownscaling}, a well-established method for rainfall downscaling that relies on a nonlinear transformation of a Gaussian random field.
A detailed description of the RainFARM approach, as well as its subsequent refinement by \citet{Terzago2018Stochasticdownscalingprecipitation} is provided in the Appendix.
In this work, we used this latest version of RainFARM when comparing with RainScaleGAN.

To ensure consistency between the data provided to the GAN during training, RainFARM was run using the slope of the power spectrum of the upscaled ERA5 training set (1940-1998).
The climatology of this data at the original resolution was used to compute the corrective weights.
Once these parameters were determined, RainFARM was applied to the upscaled (\num{2}\si{\degree}$\times$\num{2}\si{\degree}) ERA5 test set to generate precipitation fields at the target resolution (\num{0.25}\si{\degree}$\times$\num{0.25}\si{\degree}).
These fields were then used as a baseline for evaluating the GAN.


\section{Results}
\label{sec:results}

\subsection{Training analysis}
\label{subsec:training_analysis}

\begin{figure}[t]
	\centering
	\includegraphics[width=\textwidth]{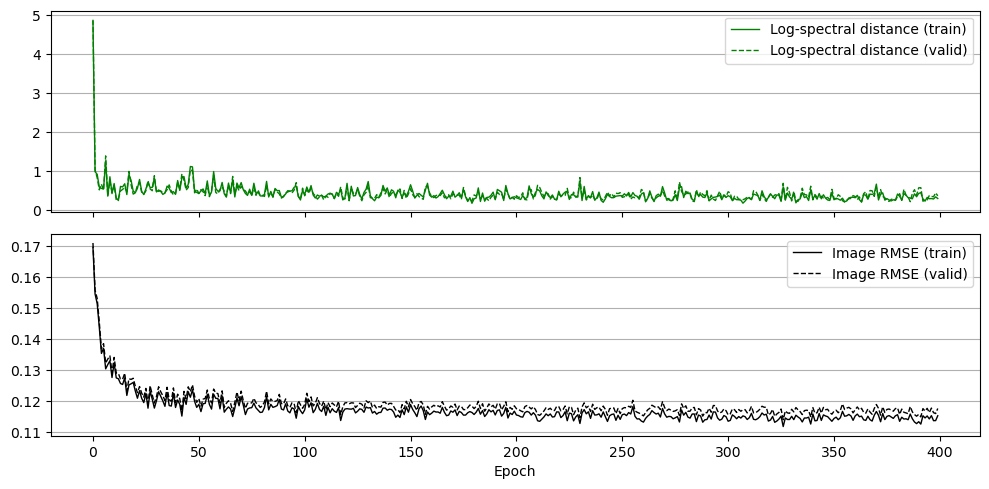}
	\caption{Evolution of image metrics throughout the GAN training process.
	(Top) Log-spectral distance between the spatial radial spectra of the generated precipitation and that of the corresponding true dataset.
	(Bottom) Root-mean-squared errors between the generated images and their corresponding true counterparts.
	The solid lines refer to the training dataset (1940-1998), while the dashed lines correspond to the validation dataset (1999-2010).}
	\label{fig:LSD_RMSE_training}
\end{figure}

\begin{figure}[t]
	\centering
	\includegraphics[width=\textwidth]{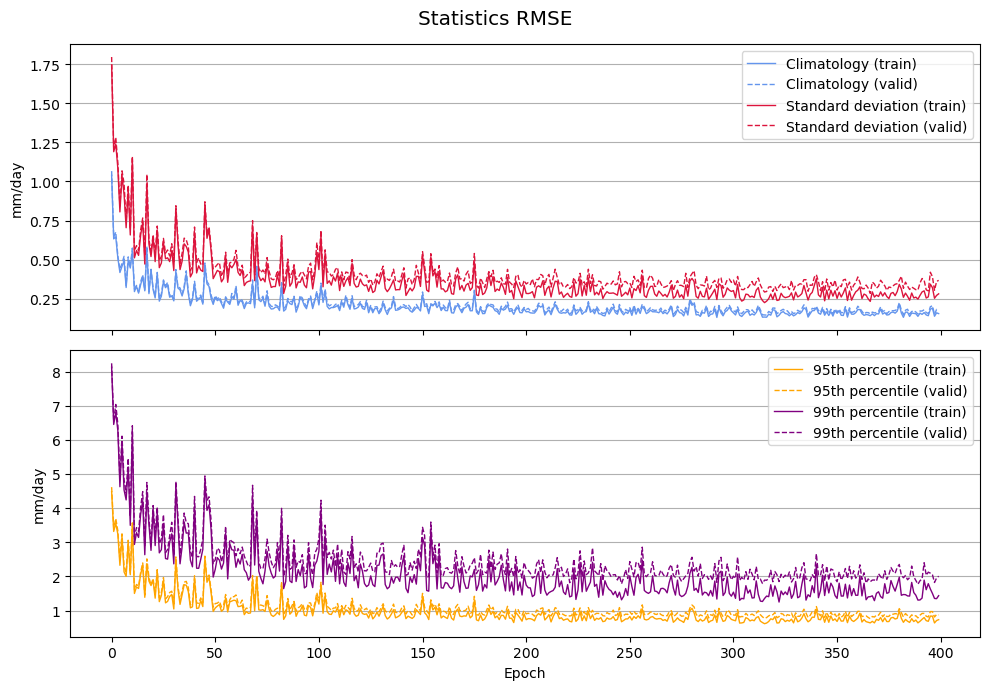}
	\caption{Evolution of statistical metrics over the GAN training: root-mean-squared errors between (top) the climatology and standard deviation, and (bottom) 95\textsuperscript{th} and 99\textsuperscript{th} percentiles of the generated dataset with respect to the corresponding true dataset.
	The solid lines refer to the training dataset (1940-1998), while the dashed lines correspond to the validation dataset (1999-2010).}
	\label{fig:statistics_RMSE_training}
\end{figure}

During training, we expect the generated dataset to progressively become similar to the training dataset.
This implies that its statistical properties will converge to those of the training dataset.
Unlike a standard GAN, RainScaleGAN - which is based on a Wasserstein GAN - has loss functions correlated with both the convergence of the generator and the quality of the generated data \citep[cf.][]{Arjovsky2017WassersteinGAN,Gulrajani2017ImprovedTrainingWasserstein}.
However, to have a more meaningful perspective on the quality of the climate delivered, we decided to rely on the set of metrics defined in Section \ref{subsec:skill_metrics} to monitor the training process.

Figure \ref{fig:LSD_RMSE_training} shows the evolution of the log-spectral distance and of the root-mean-squared error between the (spectra of) the generated precipitation fields and their corresponding true counterparts, throughout RainScaleGAN's training process.
Similarly, Figure \ref{fig:statistics_RMSE_training} shows the evolution of the root-mean-squared errors for climatology, standard deviation and 95\textsuperscript{th} and 99\textsuperscript{th} percentiles, with respect to the corresponding quantities for the true datasets.
In detail, the calculation of these metrics proceeded as follows:
\begin{itemize}
	\item At the end of each training epoch, the generator, with parameters fixed at the last update, was used to reconstruct both the training set (1940-1998) and the validation set (1999-2010), from the corresponding ERA5 upscaled versions.
	\item We computed the root-mean-squared errors between the generated datasets and their ground truth ERA5 counterparts, for both the training and the validation sets.
	The data used in this computation is bounded within the range $[-1,1]$, originating from a dataset subjected to the pre-processing operations outlined in Sec.~\ref{subsec:data_source_processing}.
	\item The generated train and validation datasets were denormalised, applying the inverse of the scaling operation (Eq.~\ref{eq:GAN_scaling}), and re-trasformed to have units of \si{\milli\meter}/day, applying a square operation.
    Please note that the same scaling factors calculated for the training set were used for scaling both the generated training and validation sets. 
	\item The climatology, standard deviation, 95\textsuperscript{th} and 99\textsuperscript{th} percentiles for both the generated and validation datasets in \si{\milli\meter}/day were calculated.
	The root-mean-squared errors of these statistics were computed, with respect to the corresponding quantities for the ERA5 datasets.
	\item The power spectra of the generated and validation datasets in \si{\milli\meter}/day were computed, from which the log-spectral distances with respect to the power spectra of the respective true datasets were derived.	
\end{itemize}

The evolution of all the metrics indicates that the generator converges during training.
A rapid improvement is observed in the first 50 epochs, followed by a slow, steady improvement, leading to a stable situation between epochs 300 and 400.
Importantly, the trend observed in metrics calculated on the validation set closely follows that of the metrics computed on the training set, which allow us to exclude the occurrence of overfitting during the training process.
After conducting several sensitivity analysis, considering the absence of evident overfitting, we empirically established that 400 training epochs is the threshold beyond which RainScaleGAN's skill does not significantly improve.

\subsection{Model selection}
\begin{figure}[!t]
	\centering
	\includegraphics[width=0.75\textwidth]{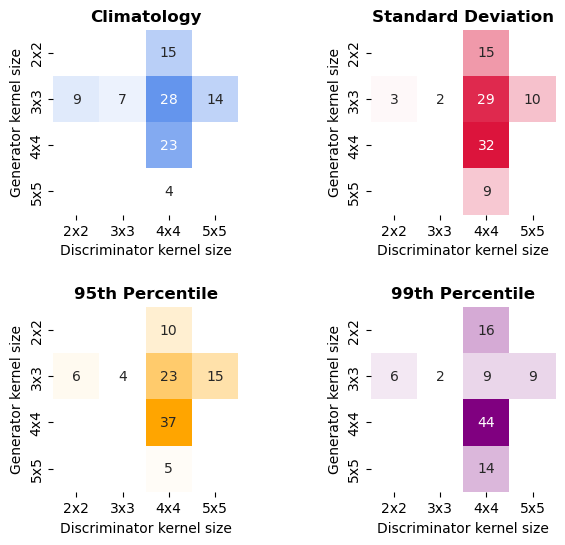}
	\caption{Selection of convolutional filter sizes. The plot shows, for each configuration and each selection metric considered, the number of training epochs in the final phase of training, after the GAN has stabilized, in which each configuration achieves the lowest root-mean-squared error for that metric compared to the other configurations.}
	\label{fig:GAN_validation}
\end{figure}

Optimising RainScaleGAN's hyperparameters is a challenging task, due to the non-monotonic behaviour of the above defined metrics during the training process (cf.~Figures \ref{fig:LSD_RMSE_training} and \ref{fig:statistics_RMSE_training}).
This difficulty can be attributed to the intrinsically stochastic nature of the model, stemming from the inclusion of a white noise source in the generator's input.
This characteristic does not compromise the quality of the GAN with respect to the downscaling task it is designed for, as a downscaling model of the type developed in this study aims at generating a (set of) possible realisation(s) of precipitation at the small scale, having statistical properties that mirror those measured in the corresponding area \citep{Rebora2006RainFARMRainfallDownscaling}.
However, this small-scale stochasticity results in the variability of the generated field, which, while being desirable in studies employing ensembles, may negatively impact grid point-wise calculated statistics, such as those considered here.
Indeed, metrics that compare a forecast and an observation (or, as in this case, a surrogate like a reanalysis) individually in each location suffer from the so-called \textit{double penalty effect} \citep{Rossa2008Overviewmethodsverification}: small errors in the placement of the forecasted precipitation result in both the penalty associated with incorrectly locating precipitation (miss) and predicting it in the wrong place (false alarm).

Instead of introducing additional metrics, we opted for a simple selection criterion.
We evaluated the model in different configurations, varying one hyperparameter at a time, and selected the configuration that performed best based on the number of epochs —after RainScaleGAN stabilized (i.e.~after 300 epochs, as specified in Section \ref{subsec:training_analysis}) - in which it achieved the lowest root-mean-squared error across a set of metrics computed for the validation set, relative to the corresponding ERA5 subset.
For this task we considered the four statistical metrics - climatology, standard deviation, 95\textsuperscript{th}, and 99\textsuperscript{th} percentiles - since they effectively evaluate the accurate reproduction of climate, which is our primary interest, and are less prone to excessive variability during the training process.
For each of the final 100 epochs in each training run within a given group of model configurations, we identified the configuration that achieved the lowest root-mean-squared error for each metric compared to all others in the group.
We separately tested the size of the convolutional filters for the generator and the discriminator, varying it between 2x2, 3x3, 4x4, and 5x5.
The results of these experiments, depicted in Figures \ref{fig:GAN_validation}, led us to choose the configuration that achieved the best scores for the most metrics considered, namely the one with 4x4 convolutional filters for both the generator and the discriminator.

\subsection{Model evaluation: analysis of a single realization}
\label{subsec:model_evaluation_deterministic}

\begin{figure}[!t]
	\centering
	\includegraphics[width=0.75\textwidth]{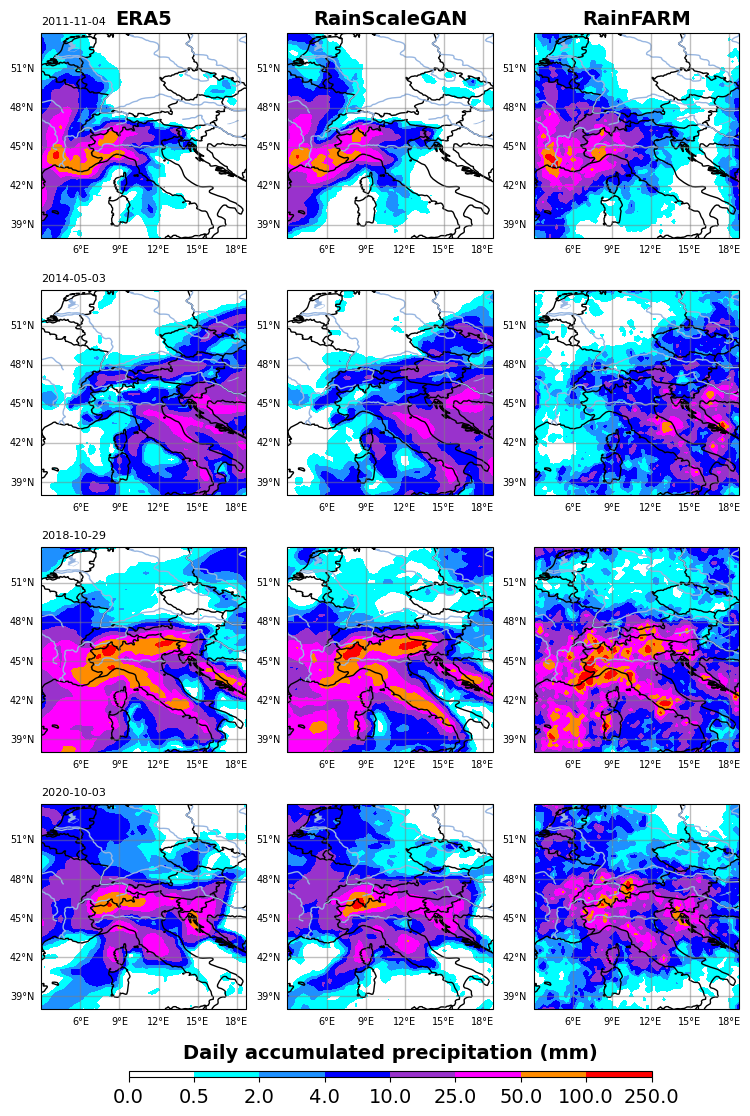}
	\caption{Comparison between the predictions generated by RainScaleGAN and the predictions of RainFARM, for four randomly selected precipitation events. The leftmost column displays the (ERA5) ground truth data.}
	\label{fig:GAN_RF_comparison}
\end{figure}

The evaluation of RainScaleGAN was conducted using the test set previously held-out for this purpose (years 2011-2022).
The preprocessing of this dataset followed the same procedure applied to the training and validation datasets, as outlined in Section \ref{subsec:data_source_processing}.
Importantly, the same scaling factors calculated for the training set were used for the scaling of the test set.
Since RainScaleGAN is trained to handle rainfall scaled with these factors, the network parameters are adjusted during training to reproduce precipitation magnitudes at each grid point that depend on this scaling.
The use of different scaling factors (calculated, for example, on the test set itself) would compromise the generator's ability to reproduce the correct amount of precipitation at each grid point, along with its capacity to extrapolate to previously unseen data.
After these operations, the optimal generator identified during the validation phase was used to downscale the entire upscaled test set, which was then denormalized and transformed back to units of \si{\milli\meter}/day.
In this way, a dataset was created that should mimic the test set extracted from the original ERA5 data.

\begin{figure}[!t]
	\centering
	\includegraphics[width=0.75\textwidth]{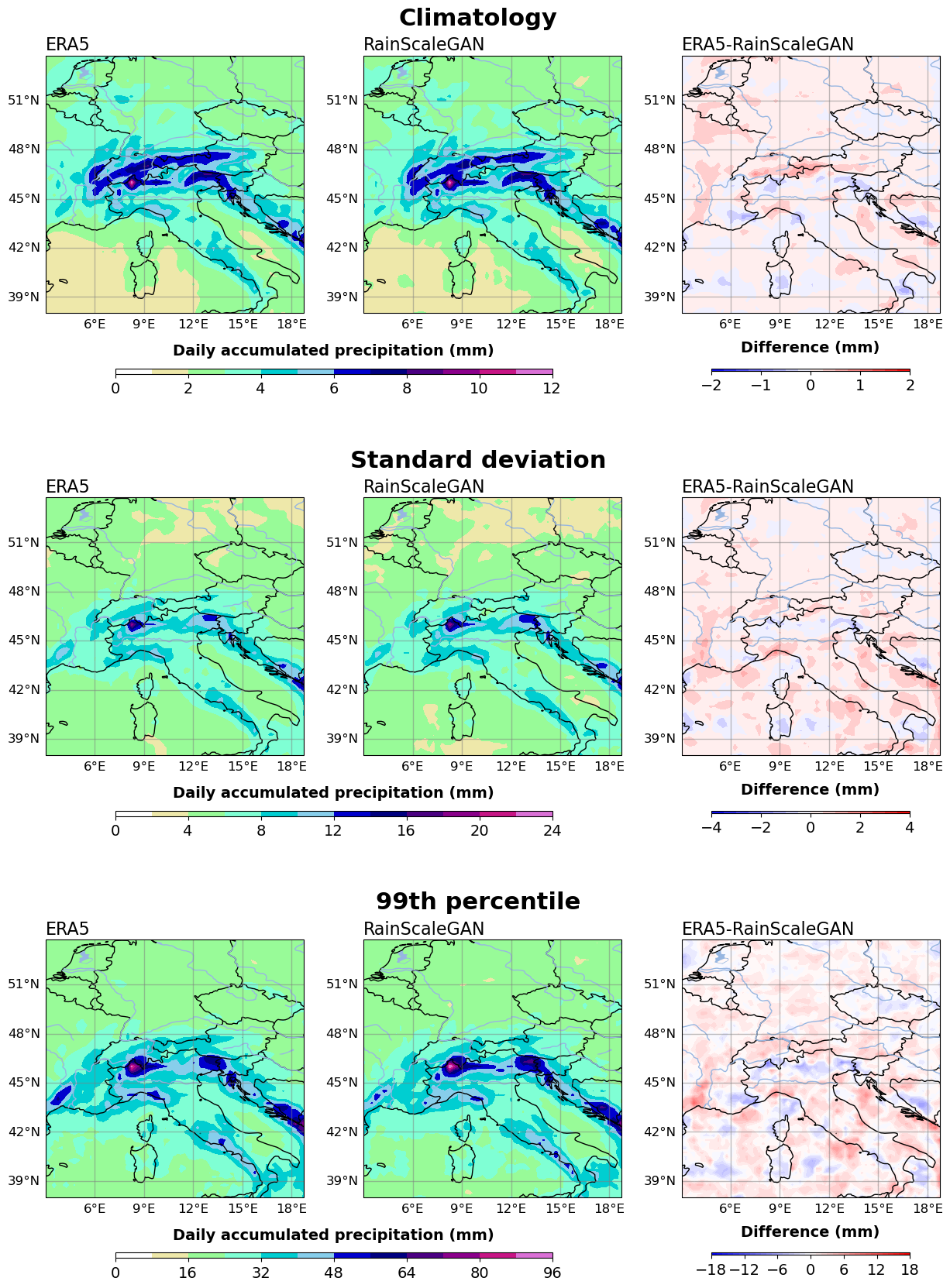}
	\caption{Maps of statistical metrics for the ERA5 ground-truth test set (2011-2022) (leftmost column), those for the test set as reconstructed by RainScaleGAN (central column), and the deviation between the corresponding metrics of the two groups (rightmost column).}
	\label{fig:statistics_diff_GAN}
\end{figure}
\begin{figure}[!t]
	\centering
	\includegraphics[width=0.75\textwidth]{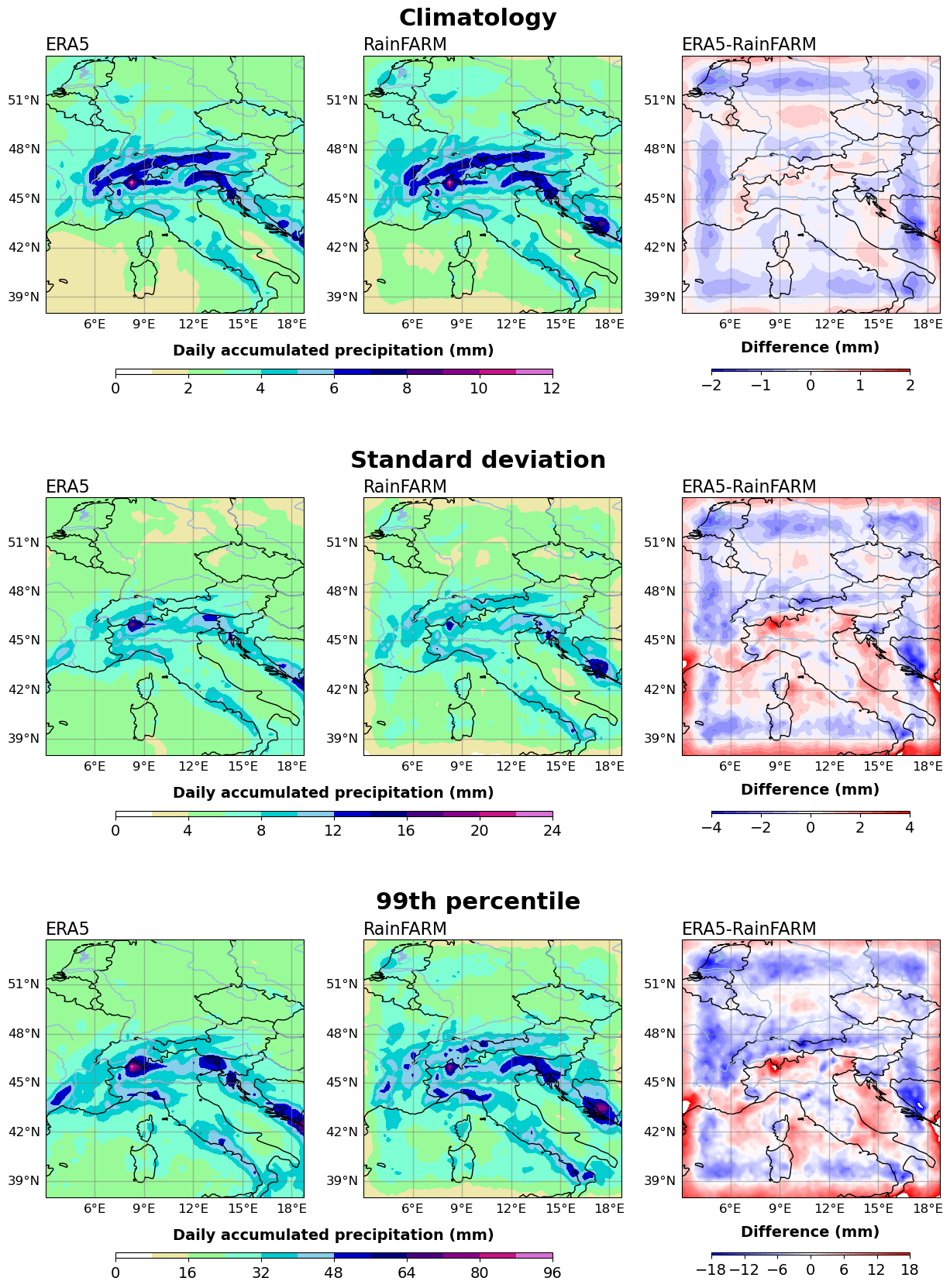}
	\caption{Same as Fig.~\ref{fig:statistics_diff_GAN} but for the test set (2011-2022) as reconstructed by RainFARM.}
	\label{fig:statistics_diff_RainFARM}
\end{figure}

Figure \ref{fig:GAN_RF_comparison} shows the comparison between the predictions generated by RainScaleGAN and those of RainFARM, for four randomly selected precipitation events, along with the corresponding ERA5 ground truth.
The analysis of the maps highlights that RainScaleGAN produces precipitation fields with more realistic details.
While both methods are effective in capturing the large-scale structure of the precipitation field, RainFARM seems constrained to reproduce fine-scale details with the same texture across all parts of the domain.
Furthermore, RainFARM's downscaling procedure, while conservative, introduces local maxima at locations distinct from those where the actual maxima are present in the original data.
In contrast RainScaleGAN, despite producing discrepancies compared to the ground truth field, appears to generate a field that is more visually consistent with the true field.
Moreover it successfully captures the position and magnitude of precipitation maxima, which is particularly desirable in the context of studies on extreme events.

The analysis of the statistical metrics for the test set confirms the superiority of RainScaleGAN.
Figure \ref{fig:statistics_diff_GAN} displays maps of the climatology, standard deviation, and 99\textsuperscript{th} percentile for the test set downscaled by RainScaleGAN.
It also includes the metrics for the ERA5 ground truth dataset, as well as the deviations of these metrics with respect to those of the GAN downscaled dataset.
Figure \ref{fig:statistics_diff_RainFARM} shows similar maps, but considering the test set reconstructed by RainFARM.
The observed edge artifacts result from the periodic boundary conditions assumed in its implementation.
In operational settings, this issue is typically mitigated by applying the downscaling procedure to a slightly larger domain than the region of interest.

Both methods reproduce climatology with sufficient accuracy, correctly capturing its spatial patterns and magnitude.
However, RainFARM benefits from incorporating climatological information for the fine-scale precipitation field during calibration, enabling it to refine and correct its predictions accordingly.
Although the hold-out test dataset is temporally distinct from the training dataset (used to define the climatology for RainFARM), and thus exhibits different statistical properties, the inclusion of climatological information related to fine-scale precipitation over the extended period corresponding to the training set enhances the RainFARM downscaling.
This approach helps capturing the spatial behaviour of precipitation, especially in regions with complex orography like the Alps, where the terrain significantly influences the spatial patterns of precipitation.
Consequently, this source of information contributes to the observed good outcome.
Conversely, the GAN does not have \textit{explicit} access to this type of information.
The accurate reproduction of climatology in areas with complex orography suggests that RainScaleGAN, by seeing during the training process examples of precipitation fields sampled from the same probability distribution it aims to reconstruct, is able to autonomously infer its statistical characteristics, including climatology.
This explains the excellent results, even without the explicit constraint provided to RainFARM.
We consider this a remarkable achievement.
For the other statistics, RainScaleGAN continues to excel in reconstruction accuracy, whereas RainFARM introduces artifacts and distortions in both the placement of local maxima and the prediction of their correct magnitudes.
Table \ref{tab:rmse_metrics} reports the root-mean-squared errors between the statistics of the datasets generated by RainScaleGAN and RainFARM, with respect to the corresponding ERA5 dataset, further confirming the above observations.

\begin{table}[t]
	\centering
	\begin{tabular}{l r r}
		\toprule
		& \multicolumn{2}{c}{RMSE (\si{\milli\meter}/day)} \\
		& RainScaleGAN & RainFARM \\
		\midrule
		Climatology                       & 0.178648 & 0.368761 \\
		Standard deviation                & 0.389718 & 1.121955 \\
		95\textsuperscript{th} percentile & 0.920116 & 2.443838 \\
		99\textsuperscript{th} percentile & 2.111995 & 5.848271 \\
		\bottomrule
	\end{tabular}
    \caption{Root-mean-squared errors for the statistical metrics of the downscaled test set, with respect to those of the ERA5 test set.}
	\label{tab:rmse_metrics}
\end{table}

\begin{figure}[!t]
	\centering
	\includegraphics[width=\columnwidth]{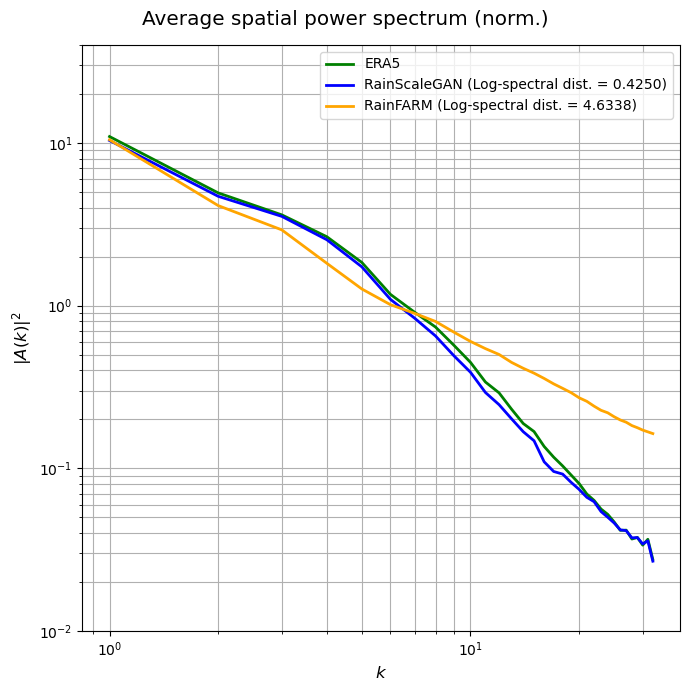}
	\caption{Time-mean radially-averaged power spectra of the downscaled test sets generated by RainScaleGAN and RainFARM, compared with the reference power spectrum of the ERA5 dataset.}
	\label{fig:GAN_RF_power_spectrum}
\end{figure}

Figure \ref{fig:GAN_RF_power_spectrum} displays the time mean of radial power spectra for the test set examples, generated by both RainScaleGAN and RainFARM, along with the corresponding reference spectrum for the ERA5 ground truth dataset. 
The figure legend also includes the log-spectral distance between the two generated power spectra and the ERA5 spectrum. 
Details for the calculation of these quantities are provided in Section \ref{subsec:skill_metrics}. 
It is evident that RainScaleGAN produces a dataset whose mean spectrum faithfully reflects that of the reference dataset, while RainFARM loses definition at small scales (high $k$), where its spectrum appears as a simple extrapolation of that at larger spatial scales.
This observation is not surprising, considering the theoretical framework of RainFARM (see the Appendix for details).

\begin{figure}[!t]
	\centering
	\includegraphics[width=\columnwidth]{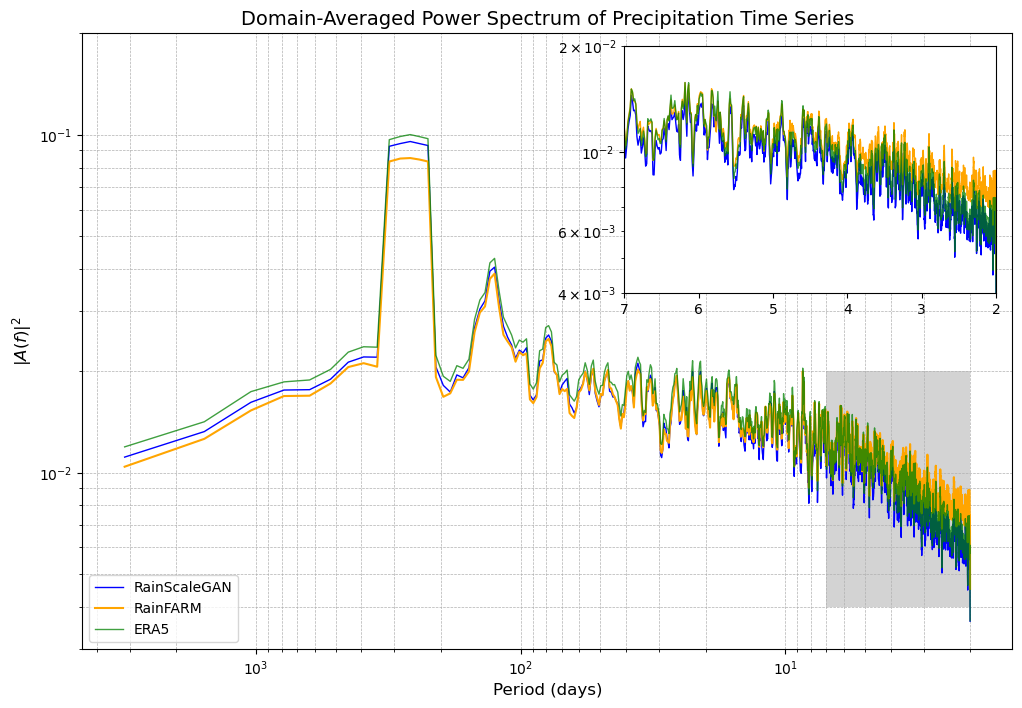}
	\caption{Domain-averaged power spectrum of the precipitation time series from the downscaled test set generated by RainScaleGAN and RainFARM, compared with the reference power spectrum from the ERA5 dataset.}
	\label{fig:GAN_RF_power_spectrum_time}
\end{figure}

RainScaleGAN does not explicitly model the temporal evolution of precipitation fields.
Consequently, the temporal structure of the downscaled outputs is expected to reflect that of the low-resolution inputs.
In other words, the temporal consistency of the generated precipitation relies solely on the model that produces the training data.
Nonetheless, it is important to evaluate whether RainScaleGAN distorts the temporal characteristics of precipitation.
To investigate this, we analyze the power spectrum of the precipitation time series, averaged over the considered domain.
Figure \ref{fig:GAN_RF_power_spectrum_time} shows the domain-averaged power spectrum of the precipitation time series from the downscaled test set generated by RainScaleGAN and RainFARM, compared to the reference power spectrum from the ERA5 dataset.
The power spectrum is computed by applying a normalized Fourier transform to the precipitation time series at each grid point within the test set (2011–2022).
The resulting spectra are then averaged over the domain.
This analysis is performed for ERA5, as well as for a single realization of the precipitation field generated by both RainScaleGAN and RainFARM.
To reduce noise, the power spectrum is smoothed using a running mean filter with a window size of 5.
A comparison of the three spectra reveals that RainFARM tends to overestimate power at short periods, suggesting an excess of variability at high frequencies.
In contrast, RainScaleGAN more closely follows the ERA5 reference spectrum, indicating a more realistic reconstruction of the temporal structure of precipitation variability.

\begin{figure}[!t]
	\centering
	\includegraphics[width=\columnwidth]{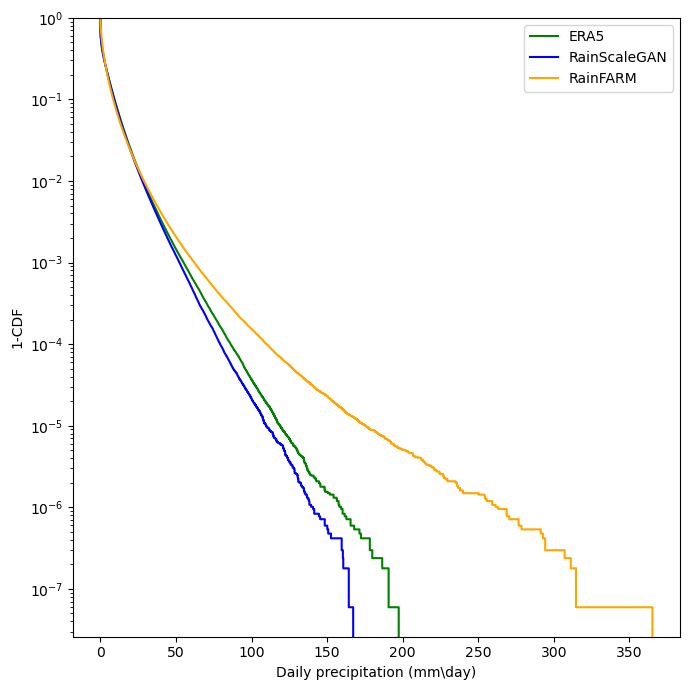}
	\caption{The complementary cumulative distribution function (CDF) of the daily accumulated total precipitation for the ERA5 test set (2011-2022), along with the corresponding functions for the test set as reconstructed by RainScaleGAN and RainFARM.}
	\label{fig:complCDF}
\end{figure}
\begin{figure}[!t]
	\centering
	\includegraphics[width=\columnwidth]{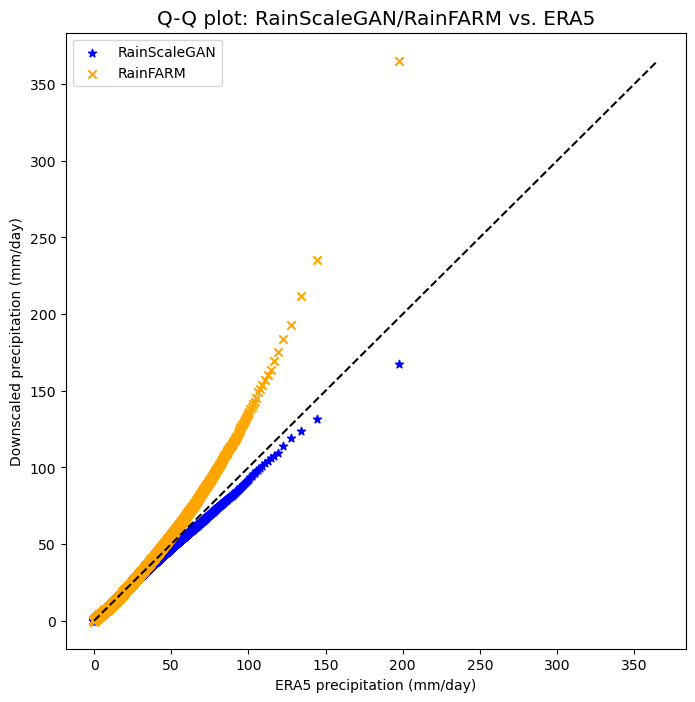}
	\caption{Q-Q plot of the precipitation probability distribution for the test dataset (2011-2022) downscaled by RainScaleGAN and RainFARM against the probability distribution of the ground-truth ERA5 test dataset.}
	\label{fig:QQplot}
\end{figure}

As an additional evaluation tool to assess the downscaling skills of RainScaleGAN, we analyzed the probability distribution of the daily accumulated total precipitation across the entire domain, over the full temporal extent of the test set.
We treated all grid points together, considering all time steps, to construct a single probability distribution.
Figure \ref{fig:complCDF} shows the complementary cumulative distribution function (CDF) of the ground truth ERA5 test dataset, together with the complementary CDFs of the test sets reconstructed by RainScaleGAN and RainFARM.
For a real-valued random variable $X$ evaluated at $x$, this quantity is defined as:
\begin{equation}
    \bar{F}_X(x) = P(X>x) = 1 - F_X(x) = 1 - \int_{-\infty}^x f_X(t) dt
\end{equation}
where $F_X$ is the CDF of $X$ and $f_X$ its probability density function.
$\bar{F}_X(x)$ represents the probability of the variable $X$ exceeding the value $x$.
In our specific case, with $X$ being the daily precipitation values from all grid points and all time steps in the test set, this has a good physical meaning, expressing the probability of a certain precipitation value being exceeded across the entire geographical region.
The functions plotted in Fig.~\ref{fig:complCDF} demonstrate that RainScaleGAN is able to accurately reconstruct the amount of precipitation over the studied domain, even though it slightly underestimate the rightmost tail of the distribution, generating slightly lower precipitation maxima.
On the other hand, RainFARM appears to overestimate the highest values of the precipitation distribution, introducing a significant number of unrealistic extreme values, exceeding the true maxima with value well above 200 \si{\milli\meter}/day.

The quantile–quantile (Q-Q) plot in Figure \ref{fig:QQplot} further confirms these observations.
It is constructed by plotting the quantiles of the precipitation probability distribution of the test dataset, as reconstructed by RainScaleGAN and RainFARM, against those of the probability distribution of the ground-truth ERA5 dataset.
Each point on the plot represents the precipitation value corresponding to a certain quantile of the probability distribution for the test dataset generated by the two downscaling models against the value of the corresponding quantile for the ground-truth ERA5 test dataset probability distribution.
This visualization highlights that while the lowest and central parts of the generated precipitation distribution are satisfactorily captured by both downscaling techniques, RainFARM introduces a positive bias in the uppermost part of the distribution.
This is evidenced by the trend of the highest quantiles, which is steeper than the bisector line.
Conversely, the trend of the quantiles of the RainScaleGAN downscaled test set closely follows the bisector line, indicating precipitation amounts more consistent with the true amounts across the entire range of values.
These observations are consistent with the earlier analysis, as all the statistical metrics of the RainFARM-reconstructed dataset showed positive biases with respect to the corresponding statistics of the ERA5 test set.

\subsection{Model evaluation: noise impact and reliability}

\begin{figure}[!t]
	\centering
	\includegraphics[width=\textwidth]{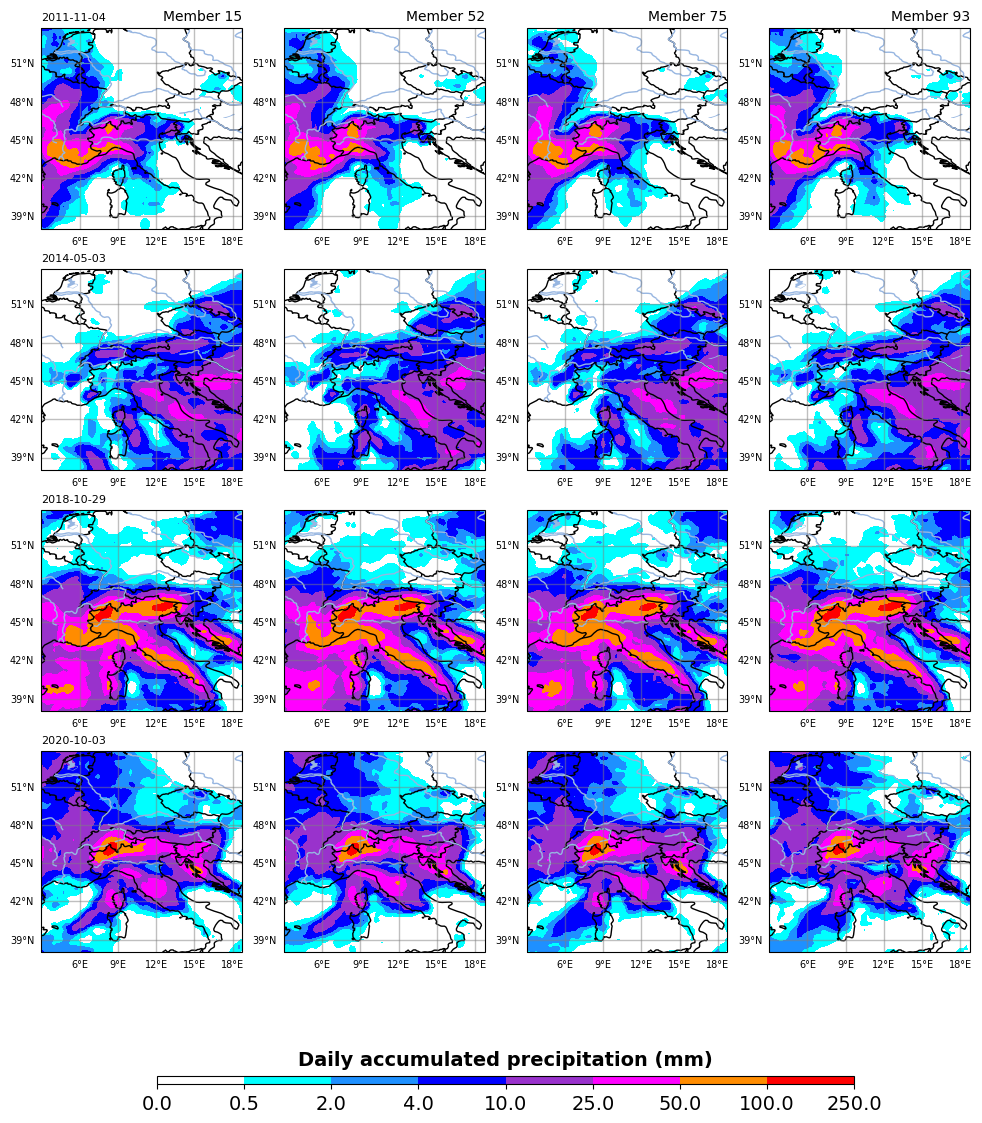}
	\caption{Four different realizations of the same set of precipitation events shown in Figure \ref{fig:GAN_RF_comparison}}
	\label{fig:RainScaleGAN_ensemble}
\end{figure}

The analysis conducted in Section \ref{subsec:model_evaluation_deterministic} was based on a single realization of the precipitation field at the target scale.
While this is important for evaluating the ability of RainScaleGAN to generate a realistic dataset with good statistical properties, there are other aspects of the output generated by the GAN that deserve further investigation.
As highlighted in Section \ref{subsec:model_architecture}, the generator includes a noise source - an array of random numbers drawn from a normal distribution with a mean of 0 and a standard deviation of 0.2.
Therefore, it is important to explore whether, and to what extent, this noise source influences the results.

To address this, we generated an ensemble of realizations of the precipitation field at the fine scale, using the test set held out for evaluation (2011–2022).
The preprocessing of this dataset followed the same steps and precautions (in particular the scaling factors) outlined in Section \ref{subsec:model_evaluation_deterministic}.
The same optimal generator identified during the validation phase was used to produce 100 realizations of the upscaled test set.
The inclusion of random noise as input to the generator ensures the stochastic nature of the fine-scale details in the output precipitation fields.
This approach allows us to investigate the impact of the noise source on RainScaleGAN output.

Figure \ref{fig:RainScaleGAN_ensemble} shows four different realizations of the same set of precipitation events considered in Figure \ref{fig:GAN_RF_comparison}.
Comparing these realizations for the same dates provides insight into the variability introduced by the noise input to the generator.
An inspection of the maps reveals that while the boundaries and positions of areas associated with precipitation events shift slightly across different realizations, RainScaleGAN appears to be self-consistent in positioning local maxima.
This consistency is important for assessing the intensity and location of heavy precipitation events.
Importantly, the large-scale spatial structure of the precipitation field is preserved, suggesting that the noise input to the generator does not distort this structure, as prescribed by the coarse-scale field being downscaled.

\begin{figure}[!t]
	\centering
	\includegraphics[width=\columnwidth]{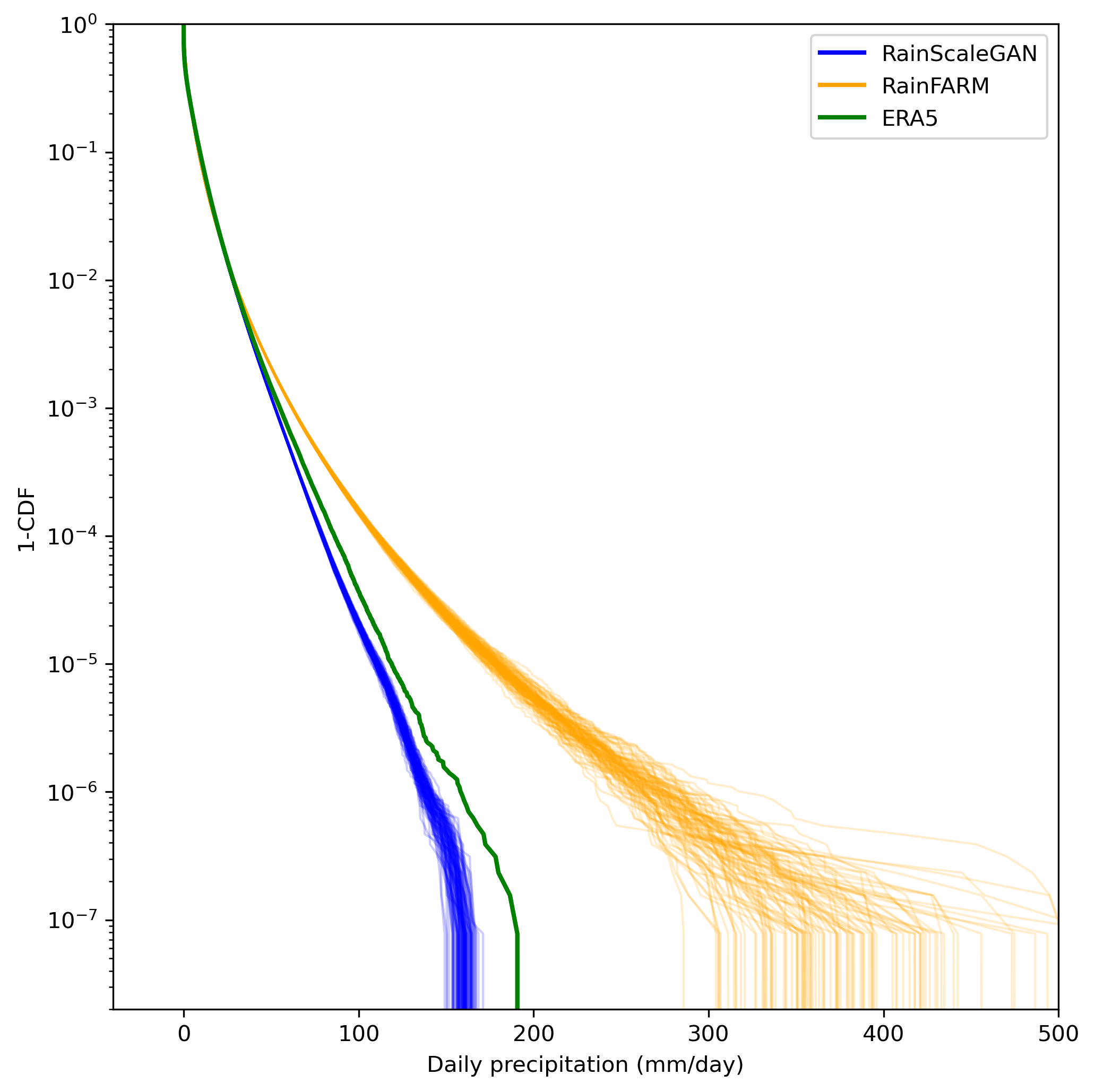}
	\caption{The complementary cumulative distribution function (CDF) of the daily accumulated total precipitation for the ERA5 test set (2011–2022), alongside the corresponding functions for 100 realizations of the test set generated by RainScaleGAN and RainFARM.}
	\label{fig:complCDF-ensemble}
\end{figure}

As in Section \ref{subsec:model_evaluation_deterministic}, to conduct a more quantitative evaluation, we considered the probability distribution for the daily accumulated total precipitation generated by RainScaleGAN by building the complementary cumulative distribution function (CDF) for the test set realizations produced by both RainScaleGAN and RainFARM.
These results were then compared with the ground truth ERA5 test set.
The corresponding plot is shown in Figure \ref{fig:complCDF-ensemble}.
This plot illustrates the probability of a certain precipitation value being exceeded across the entire considered domain.
By inspecting it, we gain insight into the spread of the precipitation maxima produced by both downscaling methods.
RainScaleGAN generates a narrower range of precipitation maxima, and, consistent with what was noted in the previous section, it slightly underestimates the rightmost tail of the distribution, yielding lower values.
In contrast, RainFARM tends to overestimate the highest values, with some members generating precipitation maxima more than double the ground truth global maximum for the accumulated precipitation within the investigated domain.
While RainScaleGAN appears to be more accurate between the two downscaling methods, its restricted range of values may be a disadvantage in studies where a good representation of variability is important.
This aspect requires further investigation, involving an assessment of the variability of results as a function of the parameters of the noise input to the generator.

Reliability diagrams are a commonly used tool for verifying probabilistic forecasts.
In these diagrams, forecasts are grouped into bins according to their predicted probability, shown on the horizontal axis, while the corresponding observed frequency is plotted on the vertical axis.
For perfect reliability, the forecast probability should match the observed frequency, resulting in points lying along the diagonal.
For a detailed explanation of their construction, interpretation, and meaning, refer to \citet{Wilks2011StatisticalMethodsAtmospheric}.

\begin{figure}[t]
	\centering
	\includegraphics[width=\textwidth]{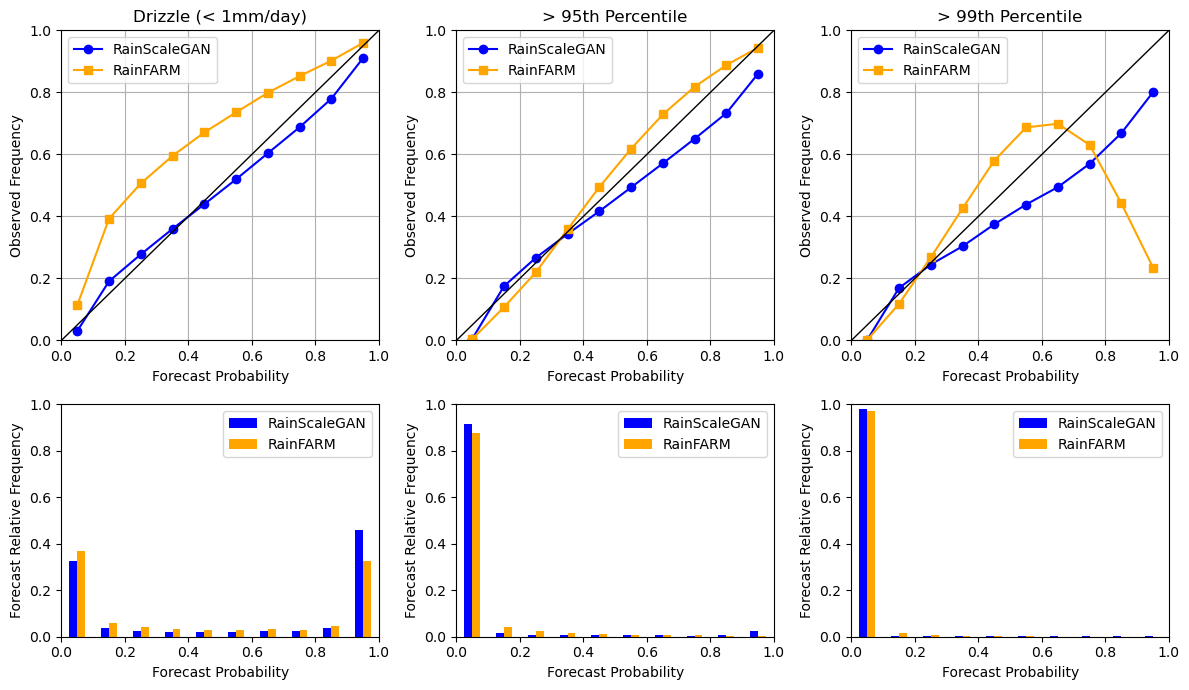}
	\caption{Domain-averaged reliability diagrams for predicted drizzle ($<$\num{1}\si{\milli\meter}/day) and extreme events ($>$\num{95}\textsuperscript{th} percentile, $>$\num{99}\textsuperscript{th} percentile) for both RainScaleGAN and RainFARM. The bottom row shows the corresponding forecast frequency histograms (sharpness diagrams), showing the relative frequency of forecast instances in each probability bin.}
	\label{fig:reliability-diagrams}
\end{figure}

Reliability diagrams are typically used to evaluate probabilistic forecasts for dichotomous events, such as rain versus no rain.
Although each realization of the precipitation field from RainScaleGAN and RainFARM can be viewed as a deterministic prediction of a continuous variable, by generating an ensemble of forecasts from the same predictor (the large-scale precipitation field to be downscaled), we can adapt this verification tool to our context.
To assess how RainScaleGAN performs across different precipitation intensities - particularly in predicting weak (drizzle) and extreme precipitation events - and to identify any potential biases, we defined three binary events:
\begin{enumerate}
    \item Total accumulated precipitation $<$\num{1}\si{\milli\meter}/day (drizzle event).
    \item Total accumulated precipitation exceeding the \num{95}\textsuperscript{th} percentile.
    \item Total accumulated precipitation exceeding the \num{99}\textsuperscript{th} percentile.
\end{enumerate}
For each time step in the test set, we evaluated the occurrence of these events at each grid point.
The thresholds for the \num{95}\textsuperscript{th} and \num{99}\textsuperscript{th} percentiles were determined from the full time series (years 2011–2022) at each grid point.
These conditions define a mask that transforms the generated dataset into a binary prediction (yes/no) for the corresponding event.
This procedure was applied separately to each ensemble member, after which the ensemble mean was computed at each time step and grid point to derive a single probabilistic prediction of event occurrence.
Using the ERA5 ground truth test set as a reference, we then computed a reliability diagram for each grid point.
Finally, we averaged these diagrams spatially across the domain to obtain the domain-averaged reliability diagram.
The results of this process are shown in Figure \ref{fig:reliability-diagrams}.
To complement the reliability diagrams, we also constructed forecast frequency histograms (sharpness diagrams), which illustrate the distribution of forecast probabilities by showing the relative frequency of instances within each probability bin.

The reliability diagrams reveal distinct behaviours for the two models.
For RainScaleGAN, reliability varies across precipitation thresholds.
For drizzle events, the model exhibits good calibration, with its reliability curve closely following the diagonal, indicating that predicted probabilities align well with observed frequencies.
For extreme precipitation, the model becomes increasingly overconfident, overestimating the occurrence of extreme events while maintaining good reliability for low forecast probabilities.
This effect is more pronounced at the \num{99}\textsuperscript{th} percentile, where deviations from the diagonal are larger.
The sharpness diagram for drizzle shows peaks at the extreme probability bins (near 0 and 1), suggesting that the model most often assigns either very low or very high probabilities.
For extreme precipitation, the distribution shifts, with a dominant peak in the first probability class (0–0.1), indicating that the model assigns low probabilities most of the time.
However, when it does predict high probabilities, it tends to be overly confident, overestimating extreme events.
This suggests that the model is generally cautious in predicting extreme precipitation but overconfident when it does.

The sharpness diagrams for RainFARM exhibit a similar pattern, with peaks at the two extreme probability bins for drizzle events and a single peak in the lowest probability class for extreme events.
These patterns indicate that RainFARM, like RainScaleGAN, assigns strong probabilities to its forecasts.
The analysis of the reliability curves reveals interesting features.
For drizzle events, RainFARM shows a dry bias, with its reliability curve lying above the diagonal.
This implies that the actual frequency of drizzle is consistently higher than the predicted probability, indicating that the model systematically underpredicts drizzle events.
For the 95\textsuperscript{th} percentile, the reliability curve is slightly steeper than the diagonal, suggesting underconfidence.
For the 99\textsuperscript{th} percentile, the curve takes a reverse U-shape, indicating that RainFARM is highly overconfident in assigning high probabilities to extreme precipitation events, resulting in overprediction.

The analysis presented in this section does not aim to provide a comprehensive evaluation of the ensemble generated by RainScaleGAN.
Instead of a full assessment — which would require the use of ensemble-specific metrics and a quantitative analysis of the sensitivity of the produced fields to the random noise input — the goal is to illustrate RainScaleGAN potential for generating consistent sets of precipitation field realizations while highlighting the role of noise in shaping the results.
In particular, Figs.~\ref{fig:RainScaleGAN_ensemble} and \ref{fig:complCDF-ensemble} emphasize the variability introduced by the generator at small spatial scales.
This variability is crucial for precipitation downscaling, as the small-scale features of precipitation events can exhibit significant differences even under the same large-scale conditions.


\section{Discussion and Conclusion}
\label{sec:discussion_conclusion}

In this study, we introduced RainScaleGAN, a Wasserstein GAN with a simple architecture specifically tailored for downscaling precipitation in climate studies.
We evaluated the performance of our model against RainFARM, a state-of-the-art stochastic downscaling method.
Our results demonstrated that the finely-tuned GAN can effectively perform downscaling in a perfect-model experiment using daily precipitation data from the ERA5 reanalysis.
The generated daily precipitation fields, when considered individually, have a more plausible appearance compared to those produced by the alternative method.
Additionally, the reconstructed dataset exhibits climate-related statistical properties that closely reflect those of the ground-truth counterpart.

The model selection part of the downscaling exercise, though challenging due to the peculiarities of the GAN training process, is important for the success of the downscaling task.
There is no guarantee that the same hyperparameters will be effective in another geographical region or even for a dataset on a different grid within the same region considered here.
Therefore, the proposed validation process must be repeated in these situations to ensure the effective application of the methodology.

The downscaling exercise was conducted between resolutions of \num{2}\si{\degree}$\times$\num{2}\si{\degree} and \num{0.25}\si{\degree}$\times$\num{0.25}\si{\degree}, covering scales typical of climate modeling.
However, since the proposed methodology does not rely on physical assumptions, there are no \textit{a priori} limitations on applying the architecture to higher spatial resolutions, targeting storm-scale resolutions relevant for weather prediction.

The spatial resolution of the input field to be downscaled, \num{2}\si{\degree}$\times$\num{2}\si{\degree}, is commonly found in climate model projections.
This naturally raises the question of RainScaleGAN generalization ability when applied to such projections.
Two key factors influence this: (1) the accuracy of the climate model being downscaled and (2) the stationarity of the transfer function implicitly learned by the generator under climate change.
The first factor arises because RainScaleGAN is not designed to correct large-scale biases, while the second relates to its assumption, as a statistical downscaling method, that the relationship between large-scale predictors and fine-scale predictands remains stationary.
Consequently, a generator trained on coarsened ERA5 data should, in principle, perform similarly to the one in this study when applied to a climate projection.
However, verifying this assumption requires further investigation.
A sensitivity analysis will be necessary to assess the model performance across different temporal periods and climate scenarios.

From the point of view of statistical downscaling, the currently adopted perfect-model setup, which is standard in the development phase, greatly simplifies the problem.
As we pointed out in Section \ref{subsec:definition_task}, it circumvents the issue related to biases between the predictor and the target dataset by adopting a unique data source.
By focusing solely on pure super-resolution, this approach can be considered the first step in constructing a proper downscaling system.
However, classical statistical downscaling methods consist of predictor-predictand relationships that are calibrated against observations.
Therefore, two distinct datasets are involved: one at the low resolution and one at the target resolution.
For instance, in an operational context, this could involve the output of a large-scale model simulation and a finer-scale observational dataset.
In such cases, the unavoidable presence of biases must be taken into account.
As a consequence, the downscaling method is also required to correct these biases, at least to some extent.
Even though it is unrealistic to expect major location biases and wrong large-scale patterns to be corrected without improving the large-scale model, correcting the amount of precipitation at the local target scale by nudging it towards observations is a realistic goal.
An extension of the methodology described here, which can also address these type of biases could have significant applications in operational contexts, where it might complement or even replace the need for computationally expensive dynamic downscaling methods based on regional models.
In this perspective, we believe our work represents an interesting first step towards this goal.

From the perspective of stochastic downscaling, which aims to generate synthetic time series for meteorological variables, RainScaleGAN shares several advantages with methods in this category and appears to outperform RainFARM, a well-established technique in the field.
Like RainFARM, RainScaleGAN relies on a single predictor - the precipitation to be downscaled - without needing additional information at either the large or small scale.
In this context, RainScaleGAN’s ability to accurately reconstruct precipitation statistics at the target resolution, particularly climatology and higher percentiles, is a notable achievement.
The quality of the statistics of the RainScaleGAN-generated precipitation dataset, which surpasses that of the RainFARM-reconstructed dataset, is particularly significant for climatological and hydrological studies.
For these applications, the accuracy of these statistics often outweighs the precision of deterministic downscaling of individual precipitation events.
Moreover, RainScaleGAN effectively captures local-scale precipitation characteristics influenced by factors such as orography, which impacts the spatial distribution over complex terrains.
Orography is a key time-invariant field considered in many downscaling techniques.
Some methods incorporate it implicitly \citep[e.g.][]{Mei2020NonparametricStatisticalTechnique}, others explicitly as a predictor \citep[e.g.][]{Harris2022GenerativeDeepLearning}, while certain approaches are entirely based on topographic information \citep[e.g.][]{Tesfa2020ExploringTopographyBasedMethods,Mital2022Downscaledhyperresolution400}.
Unlike these techniques, RainScaleGAN achieves realistic precipitation downscaling without requiring an orographic input.
Another strength stemming from the adoption of a single-predictor framework is the potential applicability of the technique to any geographical region, regardless of complex orographic features or land-sea boundaries.
Since the model does not rely on explicit geographic or topographic data, its primary limitation in this contest is the quality of the precipitation dataset used for training.

Aiming to devise a stochastic downscaling method justifies the adoption of a conditional GAN (cGAN), a deep learning framework with a higher level of complexity compared to a deterministic neural network.
The primary motivation for using a cGAN is its ability to capture the inherent stochasticity of fine-scale precipitation patterns.
Unlike deterministic models, which provide a single best-guess estimate, a cGAN can generate multiple plausible downscaled realizations, thus better capturing the variability arising from unresolved subgrid processes.
This is particularly relevant for precipitation downscaling, where small-scale features can show significant differences, even under the same large-scale conditions.
While generating an ensemble was not the primary goal of this study, we leveraged this capability to construct the reliability diagrams and demonstrate how the noise input to the generator influences the individual downscaled precipitation fields, as well as the probability distribution of daily accumulated total precipitation (cf.~Figs.~\ref{fig:RainScaleGAN_ensemble} and \ref{fig:complCDF-ensemble}).
The intention was not to conduct a full-fledged ensemble-based study, but rather to illustrate the potential of using a stochastic model to generate ensembles at a low computational cost.

The future prospects of this work include testing RainScaleGAN in realistic use cases to verify its effectiveness in bias correction, as discussed above.
Another aspect to be explored is the flexibility of RainScaleGAN in its application to different spatial scales, up to the storm scale.
In connection with this, quantifying the maximum downscaling factor - the ratio between the spatial resolution of the predictor and the target dataset - is also to be investigated.
Additionally, examining the performance of the model across various geographical domains will be essential to assess its robustness in different regions.
These investigations will contribute to a comprehensive characterisation of the proposed downscaling technique, assessing its potential applicability as a complement or alternative to dynamical downscaling methods.
Another aspect worth further analysis is the effect of the noise source in the generator input.
The capability to generate an indefinite number of precipitation fields at the target scale, all compatible with the large-scale structure prescribed by the coarse-resolution field, paves the way for studies leveraging ensembles.
This holds significant potential for estimating errors in model predictions and offers numerous advantages for climate projections and scenarios, enabling a cost-effective generation of high-resolution precipitation field ensembles.
Finally, investigating the extensibility of the proposed method to other meteorological variables, such as temperature, or more intriguingly, wind and humidity, is another prospect that deserves exploration.

\clearpage
\vspace{1em}
\noindent \textbf{\small Acknowledgments}  
\vspace{1em}

\noindent {\small
This paper is based on Chapter 3 of the PhD thesis ``Exploring Deep Learning-Based Approaches for Precipitation Downscaling'', presented by M.~Iotti at the University of Bologna in June 2024.
M.~Iotti sincerely thanks the thesis reviewers, Profs. R.~Buizza and C.~Pasquero, for their constructive feedback and valuable insights.
He also acknowledges support for his position from the ICSC High-Performance Computing, Big Data, and Quantum Computing Research Centre.
\citet{Hersbach2023ERA5hourlydata} was downloaded from the \citet{CopernicusClimateChangeServiceC3S2023ERA5hourlydata}.
The results contain modified Copernicus Climate Change Service information.
Neither the European Commission nor ECMWF is responsible for any use that may be made of the Copernicus information or data it contains.
The authors acknowledge the CINECA award under the ISCRA initiative, for the availability of high performance computing resources and support.
}

\vspace{1em}
\noindent \textbf{\small Data availability statement}  
\vspace{1em}

\noindent {\small
The code used for training, validation, and testing of RainScaleGAN can be found in the GitHub repository: \url{https://github.com/MclTTI/RainScaleGAN}.
}

\section*{Appendix: Description of the RainFARM procedure}

The RainFARM approach can be summarised in the following steps:
\begin{itemize}
	\item The spatial power spectrum of the low-resolution precipitation field $P$ to be downscaled is computed (the procedure can be extended to the temporal component of the precipitation field, which we neglect in this study).
	\item The spectrum is extrapolated to the small unresolved scales, assuming that it approximately follows a power law.
	\item A Fourier spectrum with random uniform distributed phases $g$ is generated, encompassing wavenumbers corresponding to unresolved scales.
	The inversion of this spectrum produces a Gaussian field defined on the small scales, which is then normalised to have unit variance.
	\item A nonlinear transformation of the small-scale Gaussian field is used to generate a synthetic precipitation field $\tilde{p}$. When using an exponential transformation $\tilde{p}$ is lognormal.
	\item $\tilde{p}$ is constrained to match the low-resolution field $P$ when aggregated to the resolved scales.
	This alignment is achieved through the definition of suitable weighting factors, to be applied to $\tilde{p}$.
\end{itemize}
This procedure is inherently stochastic, as varying the phases of the Fourier spectrum $g$ results in small-scale variations of the outcomes.

\citet{Terzago2018Stochasticdownscalingprecipitation} made an additional refinement to the RainFARM procedure, introducing a method to obtain more realistic fine-scale patterns of precipitation.
The goal of this adjustment is to enhance the applicability of RainFARM in climatological and hydrological applications.
Additionally, it aims to improve its ability to capture extreme events, particularly in regions characterised by complex orography.
The method relies on the availability of a fine-scale precipitation climatology, from which corrective weights for the downscaled field are derived.
The precipitation datasets downscaled with this enhanced version of RainFARM exhibit significant improvements in climatology, featuring a greater presence of fine-scale details not obtainable with the standard version, as well as enhancements in the spatial detail, placement, and magnitude of extreme values.

\bibliography{RainscaleGAN-references.bib}

\begin{thebibliography}{38}
\providecommand{\natexlab}[1]{#1}
\providecommand{\url}[1]{\texttt{#1}}
\renewcommand{\UrlFont}{\rmfamily}
\providecommand{\urlprefix}{URL }
\expandafter\ifx\csname urlstyle\endcsname\relax
  \providecommand{\doi}[1]{https://doi.org/\discretionary{}{}{}#1}\else
  \providecommand{\doi}{https://doi.org/\discretionary{}{}{}\begingroup \urlstyle{rm}\Url}\fi
\providecommand{\eprint}[2][]{\url{#2}}

\bibitem[{Annau et~al.(2023)Annau, Cannon,, and Monahan}]{Annau2023AlgorithmicHallucinationsNearSurface}
Annau, N.~J., A.~J. Cannon, and A.~H. Monahan, 2023: Algorithmic {{Hallucinations}} of {{Near-Surface Winds}}: {{Statistical Downscaling}} with {{Generative Adversarial Networks}} to {{Convection-Permitting Scales}}. \textit{Artificial Intelligence for the Earth Systems}, \textbf{2~(4)}, \doi{10.1175/AIES-D-23-0015.1}.

\bibitem[{Arjovsky et~al.(2017)Arjovsky, Chintala,, and Bottou}]{Arjovsky2017WassersteinGAN}
Arjovsky, M., S.~Chintala, and L.~Bottou, 2017: Wasserstein {{GAN}}. \textit{arXiv preprint, arXiv:1701.07875}, \doi{10.48550/arXiv.1701.07875}.

\bibitem[{Copernicus Climate Change Service~(C3S)(2023)}]{CopernicusClimateChangeServiceC3S2023ERA5hourlydata}
Copernicus Climate Change Service~(C3S), ., 2023: {{ERA5}} hourly data on single levels from 1940 to present. Copernicus Climate Change Service (C3S) Climate Data Store (CDS), \doi{10.24381/cds.adbb2d47}.

\bibitem[{Dibike and Coulibaly(2005)Dibike, and Coulibaly}]{Dibike2005Hydrologicimpactclimate}
Dibike, Y.~B., and P.~Coulibaly, 2005: Hydrologic impact of climate change in the {{Saguenay}} watershed: Comparison of downscaling methods and hydrologic models. \textit{Journal of Hydrology}, \textbf{307~(1)}, 145--163, \doi{10.1016/j.jhydrol.2004.10.012}.

\bibitem[{D'Onofrio et~al.(2014)D'Onofrio, Palazzi, von Hardenberg, Provenzale,, and Calmanti}]{DOnofrio2014StochasticRainfallDownscaling}
D'Onofrio, D., E.~Palazzi, J.~von Hardenberg, A.~Provenzale, and S.~Calmanti, 2014: Stochastic {{Rainfall Downscaling}} of {{Climate Models}}. \textit{Journal of Hydrometeorology}, \textbf{15~(2)}, 830--843, \doi{10.1175/JHM-D-13-096.1}.

\bibitem[{Ferraris et~al.(2003)Ferraris, Gabellani, Rebora,, and Provenzale}]{Ferraris2003comparisonstochasticmodels}
Ferraris, L., S.~Gabellani, N.~Rebora, and A.~Provenzale, 2003: A comparison of stochastic models for spatial rainfall downscaling. \textit{Water Resources Research}, \textbf{39~(12)}, \doi{10.1029/2003WR002504}.

\bibitem[{Feser et~al.(2011)Feser, Rockel, von Storch, Winterfeldt,, and Zahn}]{Feser2011RegionalClimateModels}
Feser, F., B.~Rockel, H.~von Storch, J.~Winterfeldt, and M.~Zahn, 2011: Regional {{Climate Models Add Value}} to {{Global Model Data}}: {{A Review}} and {{Selected Examples}}. \textit{Bulletin of the American Meteorological Society}, \textbf{92~(9)}, 1181--1192, \doi{10.1175/2011BAMS3061.1}.

\bibitem[{Goodfellow et~al.(2016)Goodfellow, Bengio,, and Courville}]{Goodfellow2016DeepLearning}
Goodfellow, I., Y.~Bengio, and A.~Courville, 2016: \textit{Deep {{Learning}}}. MIT Press.

\bibitem[{Goodfellow et~al.(2014)Goodfellow, {Pouget-Abadie}, Mirza, Xu, {Warde-Farley}, Ozair, Courville,, and Bengio}]{Goodfellow2014GenerativeAdversarialNets}
Goodfellow, I., J.~{Pouget-Abadie}, M.~Mirza, B.~Xu, D.~{Warde-Farley}, S.~Ozair, A.~Courville, and Y.~Bengio, 2014: Generative {{Adversarial Nets}}. \textit{Advances in {{Neural Information Processing Systems}}}, Curran Associates, Inc., Vol.~27.

\bibitem[{Goodfellow et~al.(2020)Goodfellow, {Pouget-Abadie}, Mirza, Xu, {Warde-Farley}, Ozair, Courville,, and Bengio}]{Goodfellow2020Generativeadversarialnetworks}
Goodfellow, I., J.~{Pouget-Abadie}, M.~Mirza, B.~Xu, D.~{Warde-Farley}, S.~Ozair, A.~Courville, and Y.~Bengio, 2020: Generative adversarial networks. \textit{Communications of the ACM}, \textbf{63~(11)}, 139--144, \doi{10.1145/3422622}.

\bibitem[{Gulrajani et~al.(2017)Gulrajani, Ahmed, Arjovsky, Dumoulin,, and Courville}]{Gulrajani2017ImprovedTrainingWasserstein}
Gulrajani, I., F.~Ahmed, M.~Arjovsky, V.~Dumoulin, and A.~Courville, 2017: Improved {{Training}} of {{Wasserstein GANs}}. \textit{arXiv preprint, arXiv:1704.00028}, \doi{10.48550/arXiv.1704.00028}.

\bibitem[{Harris et~al.(2022)Harris, McRae, Chantry, Dueben,, and Palmer}]{Harris2022GenerativeDeepLearning}
Harris, L., A.~T.~T. McRae, M.~Chantry, P.~D. Dueben, and T.~N. Palmer, 2022: A {{Generative Deep Learning Approach}} to {{Stochastic Downscaling}} of {{Precipitation Forecasts}}. \textit{Journal of Advances in Modeling Earth Systems}, \textbf{14~(10)}, e2022MS003\,120, \doi{10.1029/2022MS003120}.

\bibitem[{Hersbach et~al.(2023)}]{Hersbach2023ERA5hourlydata}
Hersbach, H., and Coauthors, 2023: {{ERA5}} hourly data on single levels from 1940 to present. Copernicus Climate Change Service (C3S) Climate Data Store (CDS), \doi{10.24381/cds.adbb2d47}.

\bibitem[{Ioffe and Szegedy(2015)Ioffe, and Szegedy}]{Ioffe2015BatchNormalizationAccelerating}
Ioffe, S., and C.~Szegedy, 2015: Batch {{Normalization}}: {{Accelerating Deep Network Training}} by {{Reducing Internal Covariate Shift}}. \textit{Proceedings of the 32nd {{International Conference}} on {{Machine Learning}}}, PMLR, 448--456.

\bibitem[{Kingma and Ba(2017)Kingma, and Ba}]{Kingma2017AdamMethodStochastic}
Kingma, D.~P., and J.~Ba, 2017: Adam: {{A Method}} for {{Stochastic Optimization}}. \textit{arXiv preprint, arXiv:1412.6980}, \doi{10.48550/arXiv.1412.6980}.

\bibitem[{Kumar et~al.(2021)Kumar, Chattopadhyay, Singh, Chaudhari, Kodari,, and Barve}]{Kumar2021Deeplearningbased}
Kumar, B., R.~Chattopadhyay, M.~Singh, N.~Chaudhari, K.~Kodari, and A.~Barve, 2021: Deep learning--based downscaling of summer monsoon rainfall data over {{Indian}} region. \textit{Theoretical and Applied Climatology}, \textbf{143~(3)}, 1145--1156, \doi{10.1007/s00704-020-03489-6}.

\bibitem[{Ledig et~al.(2017)}]{Ledig2017PhotoRealisticSingleImage}
Ledig, C., and Coauthors, 2017: Photo-{{Realistic Single Image Super-Resolution Using}} a {{Generative Adversarial Network}}. \textit{2017 {{IEEE Conference}} on {{Computer Vision}} and {{Pattern Recognition}} ({{CVPR}})}, 105--114, \doi{10.1109/CVPR.2017.19}.

\bibitem[{Leinonen et~al.(2021)Leinonen, Nerini,, and Berne}]{Leinonen2021StochasticSuperResolutionDownscaling}
Leinonen, J., D.~Nerini, and A.~Berne, 2021: Stochastic {{Super-Resolution}} for {{Downscaling Time-Evolving Atmospheric Fields With}} a {{Generative Adversarial Network}}. \textit{IEEE Transactions on Geoscience and Remote Sensing}, \textbf{59~(9)}, 7211--7223, \doi{10.1109/TGRS.2020.3032790}.

\bibitem[{Maraun et~al.(2010)}]{Maraun2010Precipitationdownscalingclimate}
Maraun, D., and Coauthors, 2010: Precipitation downscaling under climate change: {{Recent}} developments to bridge the gap between dynamical models and the end user. \textit{Reviews of Geophysics}, \textbf{48~(3)}, \doi{10.1029/2009RG000314}.

\bibitem[{Mei et~al.(2020)Mei, Maggioni, Houser, Xue,, and Rouf}]{Mei2020NonparametricStatisticalTechnique}
Mei, Y., V.~Maggioni, P.~Houser, Y.~Xue, and T.~Rouf, 2020: A {{Nonparametric Statistical Technique}} for {{Spatial Downscaling}} of {{Precipitation Over High Mountain Asia}}. \textit{Water Resources Research}, \textbf{56~(11)}, e2020WR027\,472, \doi{10.1029/2020WR027472}.

\bibitem[{Mirza and Osindero(2014)Mirza, and Osindero}]{Mirza2014ConditionalGenerativeAdversarial}
Mirza, M., and S.~Osindero, 2014: Conditional {{Generative Adversarial Nets}}. \textit{arXiv preprint, arXiv:1411.1784}, \doi{10.48550/arXiv.1411.1784}.

\bibitem[{Mital et~al.(2022)Mital, Dwivedi, Brown,, and Steefel}]{Mital2022Downscaledhyperresolution400}
Mital, U., D.~Dwivedi, J.~B. Brown, and C.~I. Steefel, 2022: Downscaled hyper-resolution (400 m) gridded datasets of daily precipitation and temperature (2008--2019) for the {{East}}--{{Taylor}} subbasin (western {{United States}}). \textit{Earth System Science Data}, \textbf{14~(11)}, 4949--4966, \doi{10.5194/essd-14-4949-2022}.

\bibitem[{Piani et~al.(2010)Piani, Haerter,, and Coppola}]{Piani2010Statisticalbiascorrection}
Piani, C., J.~O. Haerter, and E.~Coppola, 2010: Statistical bias correction for daily precipitation in regional climate models over {{Europe}}. \textit{Theoretical and Applied Climatology}, \textbf{99~(1)}, 187--192, \doi{10.1007/s00704-009-0134-9}.

\bibitem[{Price and Rasp(2022)Price, and Rasp}]{Price2022Increasingaccuracyresolution}
Price, I., and S.~Rasp, 2022: Increasing the accuracy and resolution of precipitation forecasts using deep generative models. \textit{Proceedings of {{The}} 25th {{International Conference}} on {{Artificial Intelligence}} and {{Statistics}}}, PMLR, 10\,555--10\,571.

\bibitem[{Ravuri et~al.(2021)}]{Ravuri2021Skilfulprecipitationnowcasting}
Ravuri, S., and Coauthors, 2021: Skilful precipitation nowcasting using deep generative models of radar. \textit{Nature}, \textbf{597~(7878)}, 672--677, \doi{10.1038/s41586-021-03854-z}.

\bibitem[{Rebora et~al.(2006)Rebora, Ferraris, von Hardenberg,, and Provenzale}]{Rebora2006RainFARMRainfallDownscaling}
Rebora, N., L.~Ferraris, J.~von Hardenberg, and A.~Provenzale, 2006: {{RainFARM}}: {{Rainfall Downscaling}} by a {{Filtered Autoregressive Model}}. \textit{Journal of Hydrometeorology}, \textbf{7~(4)}, 724--738, \doi{10.1175/JHM517.1}.

\bibitem[{Reichstein et~al.(2019)Reichstein, {Camps-Valls}, Stevens, Jung, Denzler, Carvalhais,, and Prabhat}]{Reichstein2019Deeplearningprocess}
Reichstein, M., G.~{Camps-Valls}, B.~Stevens, M.~Jung, J.~Denzler, N.~Carvalhais, and Prabhat, 2019: Deep learning and process understanding for data-driven {{Earth}} system science. \textit{Nature}, \textbf{566~(7743)}, 195--204, \doi{10.1038/s41586-019-0912-1}.

\bibitem[{Rossa et~al.(2008)Rossa, Nurmi,, and Ebert}]{Rossa2008Overviewmethodsverification}
Rossa, A., P.~Nurmi, and E.~Ebert, 2008: Overview of methods for the verification of quantitative precipitation forecasts. \textit{Precipitation: {{Advances}} in {{Measurement}}, {{Estimation}} and {{Prediction}}}, S.~Michaelides, Ed., Springer, Berlin, Heidelberg, 419--452, \doi{10.1007/978-3-540-77655-0_16}.

\bibitem[{Rummukainen(1997)}]{Rummukainen1997Methodsstatisticaldownscaling}
Rummukainen, M., 1997: Methods for statistical downscaling of {{GCM}} simulations. Tech. Rep. RMK 80, {Swedish Meteorological and Hydrological Institute}.

\bibitem[{Rummukainen(2010)}]{Rummukainen2010Stateoftheartregionalclimate}
Rummukainen, M., 2010: State-of-the-art with regional climate models. \textit{WIREs Climate Change}, \textbf{1~(1)}, 82--96, \doi{10.1002/wcc.8}.

\bibitem[{Schultz et~al.(2021)Schultz, Betancourt, Gong, Kleinert, Langguth, Leufen, Mozaffari,, and Stadtler}]{Schultz2021Candeeplearning}
Schultz, M.~G., C.~Betancourt, B.~Gong, F.~Kleinert, M.~Langguth, L.~H. Leufen, A.~Mozaffari, and S.~Stadtler, 2021: Can deep learning beat numerical weather prediction? \textit{Philosophical Transactions of the Royal Society A: Mathematical, Physical and Engineering Sciences}, \textbf{379~(2194)}, 20200\,097, \doi{10.1098/rsta.2020.0097}.

\bibitem[{Sha et~al.(2020)Sha, Ii, West,, and Stull}]{Sha2020DeepLearningBasedGriddedDownscaling}
Sha, Y., D.~J.~G. Ii, G.~West, and R.~Stull, 2020: Deep-{{Learning-Based Gridded Downscaling}} of {{Surface Meteorological Variables}} in {{Complex Terrain}}. {{Part II}}: {{Daily Precipitation}}. \textit{Journal of Applied Meteorology and Climatology}, \textbf{59~(12)}, 2075--2092, \doi{10.1175/JAMC-D-20-0058.1}.

\bibitem[{Terzago et~al.(2018)Terzago, Palazzi,, and {von Hardenberg}}]{Terzago2018Stochasticdownscalingprecipitation}
Terzago, S., E.~Palazzi, and J.~{von Hardenberg}, 2018: Stochastic downscaling of precipitation in complex orography: A simple method to reproduce a realistic fine-scale climatology. \textit{Natural Hazards and Earth System Sciences}, \textbf{18~(11)}, 2825--2840, \doi{10.5194/nhess-18-2825-2018}.

\bibitem[{Tesfa et~al.(2020)Tesfa, Leung,, and Ghan}]{Tesfa2020ExploringTopographyBasedMethods}
Tesfa, T.~K., L.~R. Leung, and S.~J. Ghan, 2020: Exploring {{Topography-Based Methods}} for {{Downscaling Subgrid Precipitation}} for {{Use}} in {{Earth System Models}}. \textit{Journal of Geophysical Research: Atmospheres}, \textbf{125~(5)}, e2019JD031\,456, \doi{10.1029/2019JD031456}.

\bibitem[{Wang et~al.(2021)Wang, Tian, Lowe, Kalin,, and Lehrter}]{Wang2021DeepLearningDaily}
Wang, F., D.~Tian, L.~Lowe, L.~Kalin, and J.~Lehrter, 2021: Deep {{Learning}} for {{Daily Precipitation}} and {{Temperature Downscaling}}. \textit{Water Resources Research}, \textbf{57~(4)}, e2020WR029\,308, \doi{10.1029/2020WR029308}.

\bibitem[{Wilby and Wigley(1997)Wilby, and Wigley}]{Wilby1997Downscalinggeneralcirculation}
Wilby, R., and T.~Wigley, 1997: Downscaling general circulation model output: A review of methods and limitations. \textit{Progress in Physical Geography: Earth and Environment}, \textbf{21~(4)}, 530--548, \doi{10.1177/030913339702100403}.

\bibitem[{Wilby et~al.(1999)Wilby, Hay,, and Leavesley}]{Wilby1999comparisondownscaledraw}
Wilby, R.~L., L.~E. Hay, and G.~H. Leavesley, 1999: A comparison of downscaled and raw {{GCM}} output: Implications for climate change scenarios in the {{San Juan River}} basin, {{Colorado}}. \textit{Journal of Hydrology}, \textbf{225~(1)}, 67--91, \doi{10.1016/S0022-1694(99)00136-5}.

\bibitem[{Wilks(2011)}]{Wilks2011StatisticalMethodsAtmospheric}
Wilks, D.~S., Ed., 2011: \textit{Statistical {{Methods}} in the {{Atmospheric Sciences}}}, International {{Geophysics}}, Vol. 100. 3rd ed., Academic Press.

\end{thebibliography}

\end{document}